\documentclass[a4paper,10pt]{article}

\usepackage[pdftex]{color,graphicx,hyperref}
\usepackage{amsmath,amssymb,amsthm}
\usepackage[margin=1in]{geometry}
\usepackage{setspace}
\usepackage{booktabs}
\usepackage[font=scriptsize,labelfont=bf,labelsep=period]{caption}
\usepackage{subcaption}
\usepackage[T1]{fontenc}
\usepackage[affil-it]{authblk}
\usepackage{multirow} 
\usepackage{booktabs} 
\usepackage{xspace}
\usepackage{longtable}
\usepackage{algpseudocode}
\usepackage{algorithm}

\usepackage{xr-hyper}
\usepackage[backend=biber,style=nature,url=false]{biblatex}
\newcommand{\be}{\begin{equation}}
\newcommand{\ee}{\end{equation}}

\newcommand{\eref}[1]{Eq.~(\ref{#1})}

\newcommand{\sref}[1]{Section~\ref{#1}}
\newcommand{\fref}[1]{Fig.~\ref{#1}}

\newcommand{\tref}[1]{Table~\ref{#1}}

\newcommand{\srefsi}[1]{SI Section~{\ref{#1}}}
\newcommand{\frefsi}[1]{SI Fig.~S{\ref{#1}}}

\title{Statistical inference of dynamical processes on networks}

\author[1,*]{Javier Ureña-Carrión}
\author[2,3,4,5]{Tiago P. Peixoto}
\author[1,6,*]{Gerardo Iñiguez}

\affil[1]{\small{Tampere Complexity Lab, Data Science Research Centre, Tampere University, FI-33720 Tampere, Finland}}
\affil[2]{\small{Center for Critical Computational Studies, Goethe University Frankfurt, Frankfurt am Main, Germany}}
\affil[3]{\small{Institute of Computer Science, Goethe University Frankfurt, Frankfurt am Main, Germany}}
\affil[4]{\small{IT:U, Linz, Austria}}
\affil[5]{\small{Department of Network and Data Science, Central European University, Vienna, Austria}}
\affil[6]{\small{Centro de Ciencias de la Complejidad, Universidad Nacional Autonóma de México, 04510 Ciudad de México, Mexico}}
\affil[*]{\small{Corresponding author email: javier.urenacarrion@tuni.fi, gerardo.iniguez@tuni.fi}}

\date{}

\begin{document}
\maketitle	

\begin{abstract}
Mechanisms of interaction in spreading models are central to our quantitative understanding of networked contagion processes, from disease transmission to opinion dynamics. Yet, while empirical data can reveal who interacts with whom, they rarely provide direct information about how interactions drive spreading, leaving the underlying mechanism to be inferred from observed dynamics, and selected among competing hypotheses. We propose a general framework for model selection in binary-state spreading processes on networks and show that asymptotic approximations in the thermodynamic limit can accurately predict inference outcomes in finite systems. By systematically exploring a broad parameter space, we characterize the detectability of six archetypal spreading mechanisms commonly used in the literature and find that accuracy generally increases in sparse networks, which are prevalent in real-world systems, and near phase transitions, such as the epidemic threshold of simple contagion processes. We further assess the prevalence of these mechanisms across a diverse set of empirical datasets, highlighting the impact of data preprocessing on model recovery. Our results show that statistical model selection can fail under common conditions and suggest new directions for overcoming these limitations.
\end{abstract}

\section*{Introduction}

Characterizing the temporal evolution of spreading processes in complex networked systems is crucial to our understanding of a wide array of natural phenomena, from pandemics and information diffusion to ecological dynamics and technology adoption~\cite{barrat_dynamical_2008,castellano_statistical_2009,vespignani_modelling_2012,porter_dynamical_2016,newman_networks_2018}. Usually implemented as nonlinear coupled dynamical systems~\cite{strogatz_exploring_2001,motter_networkcontrology_2015,liu_control_2016}, stochastic processes~\cite{kampen_stochastic_2007}, or (non-)Markovian dynamics~\cite{starnini_equivalence_2017}, models of dynamical processes on networks show how underlying patterns of interactions lead to emergent collective behavior. They provide mechanistic explanations for the rise of biological epidemics~\cite{pastor-satorras_epidemic_2015,nowzari_analysis_2016,kiss_mathematics_2017}, and the spread of behaviors, norms, and ideas via opinion formation~\cite{yasseri_opinion_2025,starnini_opinion_2025}, social contagion~\cite{granovetter_threshold_1978,watts_simple_2002,ruan_kinetics_2015} and cooperation~\cite{axelrod_complexity_1997}. Models of dynamics on networks also predict the existence of global states of synchronization in biochemical oscillators~\cite{haken_chemical_1984,arenas_synchronization_2008} and brain neural activity~\cite{honey_predicting_2009,ashwin_mathematical_2016}, as well as the dynamics of network flow (traffic, supply chains, information transfer)~\cite{bressan_flows_2014}, the sudden appearance of infrastructure blackouts and cascading bank defaults~\cite{haldane_systemic_2011}, and the stability of entire ecosystems~\cite{may_stability_2019}. Such varied phenomena lead to relatively similar mathematical models~\cite{goffman_generalization_1964} describing how a large number of units and their couplings change across space and time~\cite{vespignani_modelling_2012}. The dynamics of elements are typically encoded in node state variables that might be binary (spins, contagions, votes), multistate (cultural attributes) or continuous (epidemic prevalence, oscillator phases) depending on the described system, all coupled via a network of arbitrary complexity incorporating weighted~\cite{barrat_architecture_2004,unicomb_threshold_2018}, multilayer~\cite{kivela_multilayer_2014,salehi_spreading_2015,unicomb_reentrant_2019}, temporal~\cite{li_fundamental_2017,unicomb_dynamics_2021}, or even adaptive~\cite{iniguez_opinion_2009,berner_adaptive_2023} interactions.

Empirical data on networked systems typically provide information about structure, i.e. who interacts with whom, but rarely about function, i.e. the dynamical rules governing the state of a node as a function of the states of its neighbors~\cite{peixoto_graphs_2026}. This latent information is often treated as a modelling assumption, yet conclusions about the system's behavior will depend strongly on its choice. When observations of the dynamics are available, a more informative alternative is to infer the interaction rules directly from structural and dynamical data. This requires specifying a space of candidate mechanisms and a statistical procedure for identifying the model best supported by the observations. Such inverse problem differs from those most commonly studied in network science~\cite{peel_statistical_2022}, which focus either on inferring generative models of network structure~\cite{peixoto_modelling_2017,peixoto_descriptive_2023} or reconstructing network structure from dynamical observations~\cite{peixoto_network_2019,peixoto_network_2025}. Here, instead, we assume the network structure is known and observations of a dynamical process are available, while the interaction rules remain hidden.


In particular, we focus on dynamics of interchangeable binary state variables, which can be parameterized as transition rates that depend on the number of neighbors at a given state~\cite{gleeson_binary-state_2013}.
Our work presents three major contributions on the inference of spreading processes on networks. First, we show that in the case of binary-state node dynamics over static networks~\cite{gleeson_high-accuracy_2011,gleeson_binary-state_2013,ruan_kinetics_2015,yasseri_opinion_2025,starnini_opinion_2025}, the expected likelihood ratio between a true and a competing model can be expressed in terms of the relative amount of observed data and the Kullback-Leibler divergence between proposed mechanisms over a dynamical range. Second, we show that analytical approximations of detectability in the limit of large system size apply to finite systems. Focusing on the most common spreading models found in the literature, we systematically explore a parsimonious family of dynamical models on networks. We identify conditions that promote or hinder model identifiability, showing that network density induces regimes of more limited accuracy, and revealing the rich yet subtle ways in which competing generative mechanisms may explain observed data in real-world complex networked systems. 
Finally, we perform model selection on a wide range of empirical datasets including cooperation experiments, social media posts, urban traffic, supply chain dynamics, and animal interactions, uncovering the time scales of data aggregation that allow for model detectability.

\section*{Results}
\subsection*{Likelihood ratios of binary-state dynamics on networks}
\label{sec:binary-states}

\begin{figure}[t]
    \centering
    \includegraphics[width=0.95\textwidth]{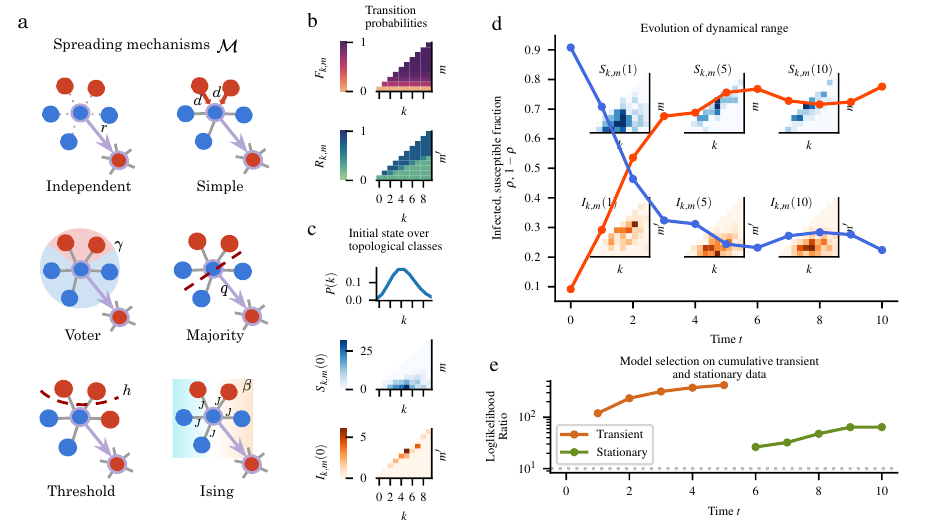}
    \caption{
    \textbf{Interplay between spreading mechanisms, network structure, and observation period affects model selection}. \textbf{(a)} Diagrams of the family $\mathcal{M}$ of spreading mechanisms considered here, each with their model parameter. Mechanisms use varying information on structural neighborhoods to infect susceptible nodes (from blue to red). \textbf{(b)} Binary-state dynamics are determined by a combination of infection and recovery mechanisms, each expressed as the probabilities $F_{k,m}$ ($R_{k,m}$) of susceptible (infected) nodes with degree $k$ and $m$ infected neighbors to switch states. We mirror the visualization of recovery matrices $m'=k-m$ to consistently display nodes with different states. We show an example dynamics with simple infections ($\theta_F=0.6$) and majority recoveries ($\theta_R=0.25$). (for all mechanisms and their parameters see \tref{tab:mechanisms}). For visualization purposes \textbf{(c)} For a given network with degree distribution $P(k)$ (here shown for a Poisson configuration model network~\cite{newman_networks_2018} with $N=250$ nodes and average degree $\langle k \rangle=5$), the initial condition of  a fraction $i_0=0.1$ of randomly selected infected nodes leads to numbers of susceptible/infected nodes $S_{k, m} (0)$ and $I_{k, m} (0)$ over $(k, m)$ classes at time $t = 0$. \textbf{(d)} As time goes by, the dynamical range of the process [defined as the time series of $S_{k, m} (t)$ and $I_{k, m} (t)$] varies across $(k, m)$ classes and leads to the fractions of infected/susceptible nodes $\rho(t)$ and $1 - \rho(t)$. \textbf{(e)} We perform model selection via the log-likelihood ratio $L_R$ over cumulative data $\mathbf{D}_F[\leq t]$ (sum of previous transient/stationary states) between the dynamics and a model alternative (here Independent, see \tref{tab:mechanisms}). Depending on whether we are in the transient or stationary state of the dynamics, this ratio might change in time.}
    \label{fig:1}
\end{figure}

Models of binary-state dynamics on networks constitute a simple yet flexible framework to explore the interplay between structure and function in complex systems~\cite{newman_networks_2018,porter_dynamical_2016,starnini_opinion_2025,yasseri_opinion_2025} (\fref{fig:1}). In these models, nodes take one of two states $X_i(t) \in \{0, 1\}$ (susceptible or infected, inactive or active, etc.) at time $t$. State changes occur stochastically based on the state $X_i(t)$ of the node and the states of its $k$ neighbors on a network, which in the simplest case is static and undirected but might be generalized to more complex scenarios~\cite{unicomb_threshold_2018, unicomb_reentrant_2019,unicomb_dynamics_2021}. The framework encompasses a wide array of dynamics on networks, such as the susceptible-infected-susceptible (SIS) model of disease spread~\cite{pastor-satorras_epidemic_2015}, the voter~\cite{holley_ergodic_1975,sood_voter_2005} and majority~\cite{de_oliveira_isotropic_1992,pereira_majority-vote_2005} models of opinion formation, threshold models of social contagion~\cite{granovetter_threshold_1978,watts_simple_2002}, and the classical Ising model of magnetic spins~\cite{krapivsky_kinetic_2010} (\fref{fig:1}a). In situations where the neighbors of a node are indistinguishable, i.e. swapping their states does not affect the dynamics, binary-state stochastic dynamics are fully characterized by the transition probabilities $F_{k, m}$ and $R_{k, m}$ with which nodes switch from state 0 to 1 and state 1 to 0, respectively, known as ``infection'' and ``recovery'' rates, where $m(t)$ is the number of ``infected'' neighbors at time $t$, for a node with $k$ neighbors~\cite{gleeson_high-accuracy_2011,gleeson_binary-state_2013}.

Transition probabilities for each class $(k, m)$ of nodes determine the model of interest and its dynamics unfolding over the network (\fref{fig:1}b).
The SIS model, for instance, has infection rate $F_{k,m}=1-(1-\theta_F)^m$ and an independent recovery rate $R_{k,m}=\theta_R$ for parameters $\theta = (\theta_F, \theta_R)$~\cite{hethcote_mathematics_2000}. The functional form of the transition rates is linear for the voter and link-update voter models~\cite{suchecki_conservation_2004}, nonlinear with an exponent $\gamma$ for its variants~\cite{schweitzer_nonlinear_2009,castellano_nonlinear_2009,abrams_modelling_2003,peralta_analytical_2018}, 
stepwise for the majority model~\cite{peralta_effect_2021,peralta_opinion_2021}, and sigmoidal for the Glauber dynamics of the Ising model~\cite{glauber_time-dependent_1963}, while the threshold model (for both absolute and relative thresholds) has a stepwise infection rate and a zero recovery rate~\cite{ruan_kinetics_2015,karsai_local_2016}. We identify six archetypic functional forms (denoted `Independent', `Simple', `Voter', `Ising', `Threshold', and `Majority', see \tref{tab:mechanisms}) and refer to them as the \textit{mechanisms} $\mathcal{M}$ driving the dynamics. These mechanisms might be different for infection and recovery (e.g., SIS model where $F_{k,m}$ and $R_{k,m}$ are probabilistically independent), or the same (e.g., voter and Ising Glauber models where transition probabilities are jointly determined). Assuming independent parameterizations for infection and recovery, we can combine all potential mechanisms for both infections $\mathbf F =\{F_{k,m}\}$ and recoveries $\mathbf R = \{R_{k,m}\}$ to define a family of models that includes standard dynamics, mixed models~\cite{dodds_limited_2013}, and otherwise unexplored combinations of $\mathbf F$ and $\mathbf R$, and where recovery probabilities are symmetric with respect to infections, $R_{k,m}=F_{k,k-m}$, i.e., where a node's state change depends on the number of neighbors with a different state.

The problem of statistically inferring a model $(\mathbf F, \mathbf R)$ of binary-state dynamics on networks from node state data corresponds to finding the values of $\mathbf F$ and $\mathbf R$ that best reproduce the observational data.
In our case, the observational data consists of the states $\mathbf X (t) = \{X_i(t)\}$ for all nodes $i\in\{1,\dots, N\}$ and time steps $t\in\{1,\dots, T\}$. For any known underlying network structure given by an adjacency matrix $\mathbf A = \{ A_{ij} \}$ with binary edges $A_{ij} \in \{0, 1 \}$, we can write the transition probabilities as
\begin{multline}
  P[\mathbf{X}(t)|\mathbf{X}(t-1), \mathbf A, \mathbf F, \mathbf R]\\
  \begin{aligned}
    &= \prod_i\left[F_{k_i,m_i(t)}^{X_i(t)}(1-F_{k_i,m_i(t)})^{1-X_i(t)}\right]^{1-X_i(t-1)}\left[(1-R_{k_i,m_i(t)})^{X_i(t)}R_{k_i,m_i(t)}^{1-X_i(t)}\right]^{X_i(t-1)}\\
    &= \prod_{k,m}F_{k,m}^{Y_{k,m}(t)}(1-F_{k,m})^{S_{k,m}(t)-Y_{k,m}(t)}R_{k,m}^{H_{k,m}(t)}(1-R_{k,m})^{I_{k,m}(t)-H_{k,m}(t)},
    \end{aligned}
    \label{eq:trprob}
  \end{multline}
with $k_i=\sum_jA_{ij}$ and $m_i(t)=\sum_jA_{ij}X_j(t)$. \eref{eq:trprob} is summarized via the quantities
\begin{subequations}
\label{eq:tr_pr_cnts}
\begin{align}
  S_{k, m}(t) &= \sum_i\delta_{k,k_i}\delta_{m,m_i(t)}[1-X_i(t)],
  \label{eq:skm_cnts}\\
  I_{k, m}(t) &= \sum_i\delta_{k,k_i}\delta_{m,m_i(t)}X_i(t),
  \label{eq:ikm_cnts}\\
  Y_{k, m}(t) &= \sum_i\delta_{k,k_i}\delta_{m,m_i(t)}X_i(t)[1-X_i(t-1)],
  \label{eq:ykm_cnts}\\
  H_{k, m}(t) &= \sum_i\delta_{k,k_i}\delta_{m,m_i(t)}[1-X_i(t)]X_i(t-1),
  \label{eq:hkm_cnts}
\end{align}
\end{subequations}
which correspond, in order, to the number of susceptible and infected nodes, and
transitions from susceptible to infected, and infected to susceptible, within
class $(k, m)$ at time $t$. Conditioned on an initial state $\mathbf X(0)$, the
total likelihood for a trajectory $\mathbf X = \{ \mathbf X (t) \}$ is then given by
\begin{equation}\label{eq:prob}
  P[\mathbf{X}|\mathbf A, \mathbf F, \mathbf R, \mathbf X(0)] = \prod_{t=1}^{T} P[\mathbf{X}(t)|\mathbf{X}(t-1), \mathbf A, \mathbf F, \mathbf R].
\end{equation}
For a given trajectory and two competing models $M_a = (\mathbf F_a,\mathbf R_a)$ and $M_b = (\mathbf F_b,\mathbf R_b)$ with no certain or impossible events on their domains (no transition probabilities one or zero), their relative posterior probabilities will be given by
\begin{equation}
  \frac{P(M_a | \mathbf X, \mathbf A)}{P(M_b | \mathbf X, \mathbf A)} = \frac{P[\mathbf{X}|\mathbf A, \mathbf F_a, \mathbf R_a, \mathbf X(0)]}{P[\mathbf{X}|\mathbf A, \mathbf F_b, \mathbf R_b, \mathbf X(0)]} \times \frac{P(M_a)}{P(M_b)},
\end{equation}
where their loglikelihood ratios can be written as
\begin{multline}\label{eq:llr}
  \log \frac{P[\mathbf{X}|\mathbf A, \mathbf F_a, \mathbf R_a, \mathbf X(0)]}{P[\mathbf{X}|\mathbf A, \mathbf F_b, \mathbf R_b, \mathbf X(0)]} = 
  \sum_{k,m} Y_{k,m} \log\left(\frac{F_{k,m}^{a}}{F_{k,m}^{b}}\right) + (S_{k,m} - Y_{k,m})\log\left(\frac{1-F_{k,m}^a}{1-F_{k,m}^b}\right) +\\
  \sum_{k,m} H_{k,m} \log\left(\frac{R_{k,m}^{a}}{R_{k,m}^{b}}\right) + (I_{k,m} - H_{k,m})\log\left(\frac{1-R_{k,m}^a}{1-R_{k,m}^b}\right),
\end{multline}
with $S_{k,m}=\sum_{t=1}^TS_{k,m}(t)$, $I_{k,m}=\sum_{t=1}^TI_{k,m}(t)$, $Y_{k,m}=\sum_{t=1}^TY_{k,m}(t)$, and $H_{k,m}=\sum_{t=1}^TH_{k,m}(t)$.

Therefore, when comparing two models, the only relevant quantities are the
time-aggregated transitions and prevalence states in \eref{eq:llr}. Besides the ability of a
model to describe the observed transitions, a determining factor for the accuracy
of model selection will be the \emph{dynamical range} observed in the
trajectory, corresponding to the different values of $(k,m)$ that are visited by
the dynamics. While inference generally improves when the amount of data is larger, transition probabilities are fixed, so the success of model selection depends on how data is allocated over the dynamical range, which results from the non-trivial interaction between  mechanisms $(\mathbf F, \mathbf R)$ over a network structure during an observation period $T$. Network structure constrains the degree classes of potential dynamical ranges [i.e. the more common degree classes and the shape of $(k,m)$ matrices], while the overall spreading dynamics can exhibit diverse macroscopic behaviors across time. These include stochastic changes over transient periods, stationary macroscopic balance between new infections and recoveries, absorbing states where node states do not change, and critical behavior close to phase transitions~\cite{barrat_dynamical_2008,dorogovtsev_critical_2008}. Within the same process, transience or stationarity can lead to differentiated dynamical ranges where candidate models are rejected to varying degrees by \eref{eq:llr} (\fref{fig:1}e). Such scenarios also impact data quantity as dynamics that continue indefinitely can produce arbitrarily large values of $S_{k, m}$ or $I_{k, m}$; yet dynamics that reach an absorbing state fast can produce few to no observations, subject to stochastic fluctuations. In the following, we will assume that all models have the same recovery mechanisms $\mathbf R$, and analyze limits to model selection of infection mechanisms $\mathbf F$ under extensive but controlled conditions for varying recovery mechanisms $\mathbf R\in\mathcal{M}$, network structures, and transient/stationary dynamics.

\begin{table}[]
    \small
    \begin{tabular}{l c c c}
    \hline
    \hline
    Mechanism $\mathbf F$, $\mathbf R$ & Parameter $\theta_F$, $\theta_R$ & Infection $F_{k,m}(\theta_F)$ & Recovery $R_{k,m}(\theta_R)$\\
    \hline
    Independent & $r$ &  $r$ & $r$ \\ 
    Simple & $d$ & $1-(1-d)^m$ & $1-(1-d)^{k-m}$ \\ 
    Voter & $\gamma$ & $(\frac{m}{k})^{\gamma}$ & $(\frac{k-m}{k})^{\gamma}$ \\ 
    Ising & $J$ (fixed $\beta$) & $\Phi\left(-2J \beta (k-2m)\right)$ & $\Phi\left(-2J \beta (2m-k)\right)$\\ 
    Threshold & $h$ (fixed $\alpha$) & $\Phi\left(-\alpha(m/k-h)\right)$ & $\Phi\left(-\alpha((k-m)/k-h)]\}^{-1}\right)$ \\
    Majority & $q$ (fixed $\alpha$)& $(1-2q)\left[\Phi(-\alpha(2m-k)\right]+q$ & $(1-2q)\left[\Phi(-\alpha(k-2m))\right]+q$\\
    \hline
    \end{tabular}
    \caption{\textbf{Mechanisms of infection and recovery in binary-state dynamics on networks}. We examine a set of six functional shapes for the rates of infection ($F_{k, m}$) and recovery ($R_{k, m}$) of binary-state spreading dynamics, each depending on a single model parameter ($\theta_F$ or $\theta_R$), node degree $k$, and number of infected neighbors $m$. The sigmoid function used for Ising, Threshold and Majority models is $\Phi(x)=\{1+e^x\}^{-1}$, the latter two having a fixed steep slope $\alpha=50$ instead of stepwise functions to ensure differentiability. Transition rates are symmetric under a $m'=k - m$ transformation, i.e. $R_{k,m}=F_{k,k-m}$. We define a spreading model $(\mathbf F, \mathbf R)$ as an independent combination of an infection mechanism $\mathbf F = \{ F_{k, m} \}$ and a recovery mechanism $\mathbf R = \{ R_{k, m} \}$ with parameters $\theta = (\theta_F, \theta_R)$. To keep one parameter per rate, the Ising mechanism is subdivided into Ising Low/High temperature (for fixed high/low values $\beta = 1, 1/10$, respectively). Our complete mechanism family $\mathcal{M}$ also includes Relative Threshold (in table) and Absolute Threshold with $F_{k,m}=\Phi(-\alpha(k-mh))$ (and corresponding $R_{k, m}$).}
    \label{tab:mechanisms}
\end{table}

\subsection*{Statistical detectability of spreading mechanisms}
\label{sec:elr}

\begin{figure}[t]
    \centering
    \includegraphics[width=\textwidth]{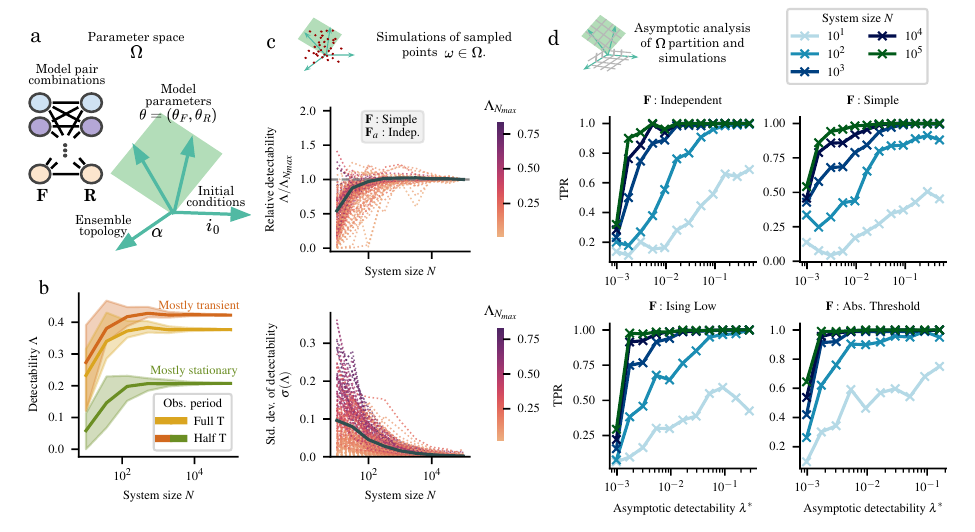}
    \caption{\textbf{Role of system size on statistical detectability of spreading dynamics.} \textbf{(a)} The parameter space $\Omega$ includes all pairwise combinations of mechanisms of infection $\mathbf F$ and recovery $\mathbf R$, their parameters $\theta = (\theta_F, \theta_R)$, a control parameter $\alpha \in \{\langle k \rangle, \sigma_k \}$ regulating network structure via its degree mean or standard deviation, and the initial fraction $i_0$ of infected nodes. \textbf{(b)} The dynamical range for some $\omega\in\Omega$ depends on the temporal evolution of node states (see \fref{fig:1}c--d), leading to a model detectability $\Lambda$ that varies with system size $N$ and temporal aggregation of data (lines). Increasing $N$ over the same observation period $T$ initially increases $\Lambda$ and decreases its variance. Detectability over transient/stationary periods may differ (red/green lines), the detectability of longer periods (full $T$, yellow line) may be lower than some shorter period (half $T$, red line) as the dynamical range also includes less detectable structural classes. \textbf{(c)} In a random sample of 250 points in parameter space ($\omega\in\Omega$) for true/alternative mechanisms $\mathbf F$ and $\mathbf F_a$, relative detectability $\Lambda/\Lambda_{N_{max}}$ increases and stabilizes with system size under a (heuristically determined) transient period. Dashed lines are averages of $\omega$ over 20 realizations for each $N$, Continuous line is average over 250 points. Colormap follows $\Lambda_{N_{max}}$, the detectability of the largest considered system ($N_{max}=10^5$). As system sizes grow, the standard deviation of detectability $\sigma(\Lambda)$ tends towards zero.  
    \textbf{(d)} We systematically analyze $\Omega$ by partitioning the space and obtaining a model's asymptotic detectability $\lambda^*$ over the partitioned space, for the same points we perform model selection in simulated data for different $N$ (each cross has 50 points with 10 simulations each). Lower $\lambda^*$ values yield lower true positive ratios (TPRs) when identifying $\mathbf F$ across system sizes, validating the suitability of $\lambda^*$ to characterize regions where model selection might fail. Panels depict TPRs for four mechanisms $\mathbf F$: independent, simple, low-temperature Ising, and absolute threshold.}
    \label{fig:fig2}
\end{figure}

Statistical model selection requires a clearly defined family of alternative model hypotheses. Given a true generative infection mechanism $\mathbf F$, we consider alternative hypotheses as the remaining non-generating mechanisms $\mathbf F_a \in \mathcal{M}_{-F}=\mathcal{M}/\{\mathbf F\}$ (see \tref{tab:mechanisms}). We define an extensive parameter space $\Omega$ that captures relevant characteristics for inference in empirical settings, including the independent product of mechanisms $(\mathbf F, \mathbf R)\in \mathcal{M} \times \mathcal{M}$ and the model parameters $\theta=(\theta_F, \theta_R)$ of each pair of infection/recovery mechanisms, structural features like system size $N$ (number of nodes) and, assuming networks sampled from the configuration model \cite{newman_networks_2018}, the average degree $\langle k \rangle$ and degree standard deviation $\sigma_k$. Overall, $\Omega = (\mathbf F, \mathbf R, \theta_F, \theta_R, \alpha, i_0)$ with $\alpha \in \{\langle k \rangle, \sigma_k \}$ the chosen control parameter for network structure and $i_0$ the initial fraction of infected nodes (\fref{fig:fig2}a). By partitioning and systematically evaluating points $\omega \in \Omega$ in parameter space, we characterize how likely we are to select a true generative model $\mathbf F$ under alternatives $\mathcal{M}_{-F}$. 
Given some point $\omega$, data quantity $S$ increases with both system size $N$ and a longer observation period $T$. We treat $N$ parametrically and distinguish between transient and stationary dynamics using the following heuristic: we simulate dynamics for at most $T_{\mathrm{max}}=100$ steps and estimate the transient as twice the relaxation time $\tau_e$ under exponential decay (for details see Materials and Methods [MM]). We assess the robustness of our results using additional heuristics (Supplementary Information [SI] \sref{sec:si:extra-heuristics}). 

For a true infection mechanism $\mathbf F$ with parameter $\theta_F$, associated data $\mathbf{D}_F=(\mathbf S, \mathbf Y)$ with $\mathbf S = \{ S_{k, m}\}$ and $\mathbf Y = \{ Y_{k, m}\}$ according to \eref{eq:tr_pr_cnts}, and an alternative mechanism $\mathbf F_a$ to explain the data, we measure the \textit{detectability} $\Lambda$ of a true mechanism (under observed data and a mechanism alternative) as a linear combination of Kullback-Leibler divergences $D_{KL}$ weighted by the observed dynamical range $S_{k, m}$ and the amount of data $S$, 
\begin{equation}
    \Lambda(\mathbf F| \mathbf{D}_F, \mathbf F_a) = \sum_{k,m}\frac{S_{k,m}}{S}D_{KL}(F_{k,m}||F_{k,m}^a),
\label{eq:elr}
\end{equation}
where $D_{KL}$ is the statistical distance between probability distributions $F_{k, m}$ and $F_{k, m}^a$. \eref{eq:elr}
is obtained by taking the expected value of the random variables $Y_{k,m}$ in the loglikelihood ratio of \eref{eq:llr}, i.e. $\mathbb{E}[Y_{k,m}]=S_{k,m}F_{k,m}$ (for derivation see MM). In this sense, a model is more detectable if data is allocated to regions of where functional forms differ more. However, the detectability of a model at point $\omega$ can vary as a function of transience/stationarity and system size (\fref{fig:fig2}b). Whereas loglikelihood ratios 
can grow infinitely with system size (which generally augments the amount of data $Y$ and $S$), we expect detectability to stabilize with system size $N$ --- subject to the observation period being fixed (e.g., transient or stationary).

Our central claim is that asymptotic approximations of detectability in the infinite size limit ($N \to \infty$) can inform model selection, even for finite systems. 
To assess the behavior of $\Lambda$ as a function of system size under transient dynamics (defined using the same heuristics), we track detectability at a point $\omega$ relative to the detectability of a very large reference system, $\Lambda_N/\Lambda_{N_{max}}$ with $N_{max}=10^5$. We sample random points from $\Omega|\mathbf F$ and run simulations on transient states of synthetic networks with varying $N$ (\fref{fig:fig2}c). On average, detectability increases and stabilizes as a function of system size for all mechanism combinations in our parameter space (all models in \frefsi{fig:si:detect_station}). 
In turn, for different realizations of the same $\omega$ the standard deviation of detectability $\sigma(\Lambda)$ decreases as systems grow (as opposed to the unnormalized detectability $S\times\Lambda$ whose variability increases with $N$, see \fref{sec:si:synth-detect}). We use these results to motivate the use of an asymptotic approximation to $\Lambda$ to systematically explore our broad parameter space $\Omega$. As detectability is dependent on the aggregation period, for completeness we include additional heuristics in \sref{sec:si:extra-heuristics}. 

Approximate Master Equations (AMEs) are high-accuracy analytical approximations for binary-state dynamics on networks explored widely in the literature~\cite{gleeson_high-accuracy_2011,gleeson_accuracy_2012,gleeson_binary-state_2013,porter_dynamical_2016,unicomb_dynamics_2021,starnini_opinion_2025,yasseri_opinion_2025}. AMEs track the average fractions of susceptible $s_{k,m}(t)$ and infected $i_{k,m}(t)$ nodes in class $(k, m)$, as coupled systems of ordinary differential equations, assuming the thermodynamic limit $N \to \infty$ and an uncorrelated, configuration-model network~\cite{newman_networks_2018} with given degree distribution $P_k$ (for explicit equations see MM). AMEs provide a natural framework for estimating the dynamical range of competing spreading mechanisms as they focus on $(k, m)$ classes, are significantly faster to compute numerically and less susceptible to stochastic fluctuations than simulations. Via AMEs we numerically estimate the cumulative ratio of susceptible nodes $\bar{s}_{k,m}(t^*)=\int_{t^*}s_{k,m}(t)dt$ over a transient period $t^*$ (computed with heuristics, see MM for details), and use the normalizing factor $\bar{s}(t^*)=\sum_{k,m}P_k\bar{s}_{k,m}(t^*)$ to arrive at the asymptotic detectability
\begin{equation}
    \lambda(\mathbf F|\mathbf{d}_{\omega}, \mathbf F_a) = \sum_{k,m}\frac{P_k\bar{s}_{k,m}}{\bar{s}}D_{KL}(F_{k,m}||F^a_{k,m}),
\label{eq:adetect}
\end{equation}
where $\mathbf{d_{\omega}}=(\mathbf{y}_{\omega}, \mathbf{s}_{\omega})$ is the observational data generated using AMEs for some point $\omega\in\Omega$. We calculate $\lambda$ for a set of alternative mechanisms $\mathcal{M}_{-F}$, and unless otherwise specified, we refer to asymptotic detectability $\lambda^*$ as the minimum over the model candidate set $\mathcal{M}_{-F
}$ (see MM for details and \sref{sec:si:comparison_synthame} for results on maximum likelihood estimates with asymptotic approximations).
To assess the extent to which asymptotic detectability $\lambda^*$ can aid in model selection, we run simulations on synthetic networks of varying size and calculate the True Positive Ratio (TPR) when identifying the true model under alternatives $\mathcal{M}_{-F}$, i.e. the true-positive fraction $TP$ of simulated samples where the true model was recovered, relative to all test including the false-negative fraction $FN$ of samples where another model in $\mathcal{M}_{-F}$ had a higher likelihood, $TPR=\frac{TP}{TP+FN}$ (\fref{fig:fig2}d). Our results show that asymptotic detectability $\lambda^*$ can systematically identify regions where model selection fails. Generally, low $\lambda^*$ implies a lower capacity to successfully perform inference. The task is consistenly more difficult in smaller systems, yet in the more extreme cases, model selection is hindered even for large system sizes (all mechanisms in \sref{sec:si:ames_add}). %

\subsection*{Detectability increases with network sparsity and near phase transitions}

\begin{figure}[t]
    \centering
    \includegraphics[width=\textwidth]{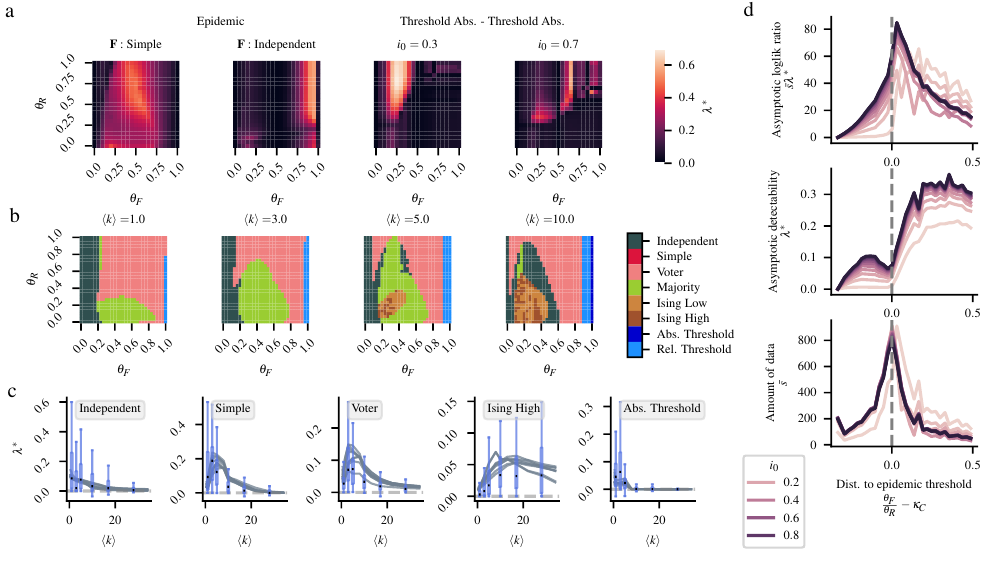}
    \caption{\textbf{The heterogeneity of asymptotic detectability $\lambda^*$ over $\Omega$ is subject to parameter regions, network density, and critical behavior.} \textbf{(a)} The asymptotic detectability $\lambda^*$ of an infection model $\mathbf F$ varies over the parameter space $\theta$ for an epidemic process (leftmost heatmaps) and competing threshold (rightmost heatmaps) dynamics on a fixed structure $\langle k \rangle=5$. Initial conditions have a large effect on the parameter region where Abs. Threshold model is detectable. \textbf{(b)} Asymptotic approximations reveal potential replacement patterns for common mechanisms. Colors represent mechanisms that most resemble Simple infections (lowest $\lambda$ in $\mathcal{M}_{-F}$) over the same parameter space $(\theta_F, \theta_R, i_0=.5)$ while increasing system density (columns). The same Simple infections can be best approximated by a wide range of mechanisms depending on the parameter region. 
    \textbf{(c)} Increasing network density decreases $\lambda^*$ for most models. Each subplot represents an infection mechanism $\mathbf F$, with $\lambda^*$ averaged per recovery mechanism $\mathbf R$ (grey lines). Violin plots represent distributions for 2500 normalized likelihood ratios ${L_R}^*/S$ on large synthetic networks $N=10^5$. Only Ising High displays a broader detectability range over degrees. \textbf{(d)} The asymptotic likelihood ratio $E^*[L_R]$ for SIS dynamics peaks after the epidemic threshold (here $\langle k \rangle=3$). This peak is largely driven by larger amounts of data $\bar{s}$ on the threshold $\kappa_C$ (bottom). The normalized detectability $\lambda^*$ has systematically lower values before and on $\kappa_C$. }
    \label{fig:fig3}
\end{figure}

We explored a wide range of dynamics on the partition of $\Omega$, an extensive parameter space where asymptotic detectability generally displays heterogeneous behavior.  
We identified salient trends in terms of structural regimes and critical phenomena; however, detectability is typically nonlinear and context-dependent rather than generalizable on every analytical axis --- say, some $\theta$ or $i_0$ (\fref{fig:fig3}). In an epidemic process on a fixed structure, detectability depends nonlinearly on both the Simple infection parameter (decreasing as $\theta_F \rightarrow 0,1$) and the Independent recovery parameter. Conversely, on the symmetric epidemic process with Independent infections and Simple recoveries, the regions of higher detectability differ when inferring the Independent infections. Similarly, in threshold models detectability is largely shaped by initial conditions. On the same structure, a parameter region with higher detectability can sharply decrease as initial conditions change. This heterogeneity extends to replacement mechanisms that with smallest asymptotic detectability $\lambda^*$ within our set of candidate mechanisms $\mathcal{M}_{-F}$, revealing a rich landscape of model alternatives for Simple infections (\fref{fig:fig3}b).The mechanisms that most resemble Simple infections include Voter, Independent, Majority, Ising, and Threshold models in varying proportions across parameter regions (additional results on \ref{sec:si:altmechs}). 

Despite the overall heterogeneity of detectability, we identified trends for network density and near critical points. 
Aggregating over non-topological parameters reveals that average asymptotic detectability $\lambda^*$ generally decreases as mean degree $\langle k \rangle$ increases (\fref{fig:fig3}c). While this occurs for all mechanisms except high-temperature Ising, there are also mechanism-specific characteristics: Independent, low-temperature Ising and Threshold models display generally decreasing detectability; whereas Simple and Voter models have peaks in detectability for different recovery mechanisms $\mathbf R$. 
When interpreting these results, a key consideration is the aggregation over all other parameters: peaks in detectability identify structural regimes where broad parameter regions are more detectable, rather than suggesting that model selection is not possible in dense networks at all (which also depends on the amount of data). This points to a link between structural regimes and detectable parameter regions, namely, as higher average degrees give nodes more infection opportunities, the parameter regions with higher detectability also become smaller. In turn, over larger spaces of possible dynamical ranges (via higher average degrees, or larger $(k,m)$ classes), the actual dynamical ranges produced by broad parameter regions are easier to approximate with alternative mechanisms. For dense networks mechanisms are more likely to be similar to others. These alternative mechanisms, however, are not always easy to systematically characterize and also include heterogeneity in alternatives (see \ref{sec:si:altmechs}).

Binary-state dynamics can exhibit non-linear behavior characterized by sharp transitions in global outcomes~\cite{dorogovtsev_critical_2008}. Critical points are associated with large sensitivity to changes on the control parameter, so it's generally assumed that inference tends to or becomes easier at the critical point. We examine SIS dynamics on ER networks, for which the epidemic threshold occurs when the actual infection rate $\theta_F/\theta_R$ equals the inverse average degree $\kappa_C={\langle k \rangle}^{-1}$. Instead of only using asymptotic detectability $\lambda^*$, we also consider asymptotic likelihood ratios $l_\mathcal{R}(\mathbf F|\mathbf F_a, \mathbf{d}_{\omega})$, which we define using data from from stochastic approximations on Eq.\ref{eq:llr} (see MM). 
Focusing on Simple infections on a fixed structure, we aggregate the average asymptotic likelihood ratio $\bar{s}\lambda^*$ over model parameters $(\theta_F, \theta_R)$, finding that it is highest closest to the epidemic threshold $\kappa_C$ (\fref{fig:fig3}d). Crucially, the largest values are located on the region where the infection process does not become extinct. This peak is driven by the interplay between the dynamical range and the amount of data $\bar{s}$ ($\lambda$'s normalizing factor in Eq.\ref{eq:adetect}). Close to $\kappa_C$ detectability is low on the on the extinction regime, with higher values after the epidemic threshold. In our results, the amount of data $\bar{s}$ is largest on the epidemic threshold itself --- a point associated with critical slowing down where infections and recoveries occur at nearly equal rates and produce more observable transitions. These results are contingent to our heuristics for transient/stationary dynamics and finite-size effects. 
However, they show that model selection is more likely to be successful at a well-known phase transition, and that this is likely to be a result of larger amounts of data rather than how that data is allocated over the dynamical range.

\subsection*{Coarse-graining determines model selection in empirical spreading data}

\begin{figure}[t]
    \centering
    \includegraphics[width=\textwidth]{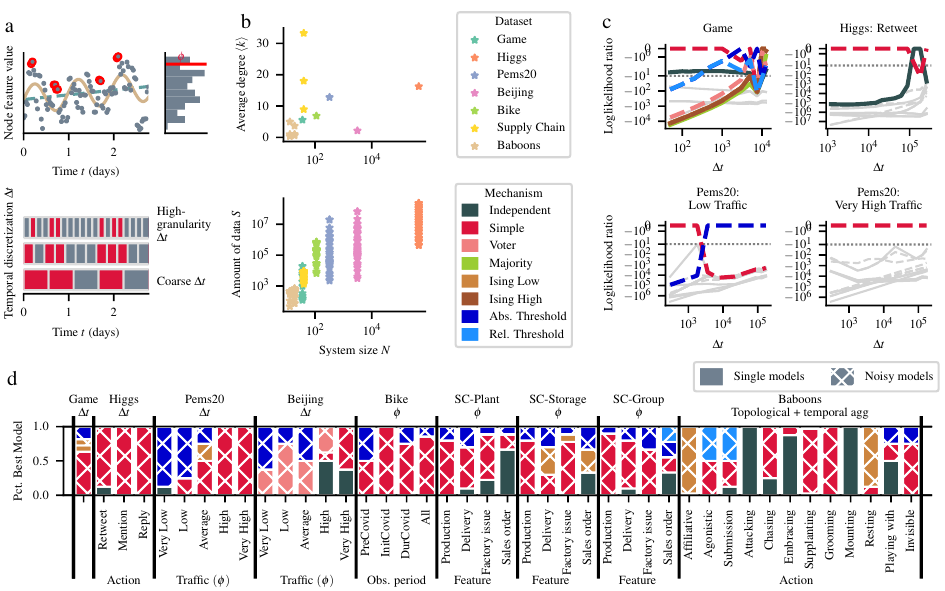}
    \caption{\textbf{Coarse-graining and preprocessing determine model selection in empirical spreading processes.} \textbf{(a)} For datasets with continuous features we deseason node-level daily/weekly patterns and trends, and binarize on the distribution of deseasoned data using a threshold $\phi$. We vary the observation window size $\Delta t$ to discretize continuous time and obtain different temporal resolutions. Both preprocessing techniques can fundamentally alter the resulting binary-state data. \textbf{(b)} We analyze seven datasets of empirical spreading phenomena. We depict average degree $\langle k \rangle$ and amount of data $S$ as a function of system size $N$. Colors represent datasets and dots different subdatasts resulting from varying preprocessing parameters. \textbf{(c)} Model selection on four subdatasets when varying temporal aggregation window $\Delta t$. We compute likelihood ratios with respect to the optimal model (valued zero). Temporal scales affect model selection: for Game, model selection is likely to fail for bigger observation windows, Higgs are mostly noisy simple infections, but resemble independent dynamics at around $\Delta t\approx1
    \text{ day}$. For Pems20, low amounts of traffic can spread as noisy simple or threshold infections depending on the temporal scale, but very high traffic is consistently simple. \textbf{(d)} For each dataset we identify a primary variable ($\Delta t$ or $\phi$, on top) and compute the relative fraction each model is selected when varying the primary variable (colors). Bar names represent subdatasets obtained from data-specific features or binarization (denoted $\phi$). The noisy versions of models dominate. While all model families appear in the In most datasets, noisy simple infections and noisy absolute threshold models dominate.  
    }
    \label{fig:fig4}
\end{figure}

We apply our framework on empirical spreading processes where both explicit network structures and temporal spreading data are available. The datasets we analyze include game-theoretic experiments on synthetic networks~\cite{kearns_behavioral_2009}, adoption of behavior on social media with underlying follower relationships~\cite{de_domenico_anatomy_2013}, traffic flow on road networks, demand on bike-sharing systems~\cite{cai_urbandata_2023}, production within large supply chains~\cite{wasi_supplygraph_2024}, and spread of baboon behavior on Bluetooth proximity networks~\cite{gelardi_detecting_2019}. All datasets have specific preprocessing requirements.
To aid in the presentation of our results we follow a general two-step pipeline, where we identify \textit{subdatasets} (e.g., resharing as different from commenting on social media),  and a \textit{primary preprocessing parameter} that controls the discretization of either time or spreading features. Preprocessing is required for potentially non-Markovian time and non-binary states. In this sense, temporal coarse-graining can capture phenomenologically district processes as, e.g., minute-by-minute social media usage might differ from day-to-day dynamics.  
Table \ref{tab:data} summarizes the structural features, subdatasets, temporal characteristics and primary parameters of our data. For example, originating from a game-theoretic study, \textit{Game} has a fixed structure and binary states by design~\cite{kearns_behavioral_2009}, leading to the observation window $\Delta t$ as the only (and primary) preprocessing parameter. \textit{Bike} has a fixed structure and bi-hourly temporal granularity for a period of a year and a half on the onset of the Covid pandemic. We analyze subdatasets of the three six-month periods before the pandemic, during its initial outbreak, and during its subsequent entrenchment. The primary parameter controls for the binarization of bike demand $\phi$, performed after deseasoning for daily and weekly patterns (\fref{fig:fig4}a). For datasets that require both temporal and feature discretizations we first perform node-level binarization with deseasoning, and subsequently discretize time (see MM and \srefsi{sec:si:preproc}). 

\begin{table}[]
\resizebox{\columnwidth}{!}{%
\begin{tabular}{@{}c|ccccc@{}}
\toprule
\textbf{Dataset} & \textbf{Structure} & \textbf{Subdatasets} & \textbf{Obs. period} & \textbf{Temp. Resolution} & \textbf{Primary parameter} \\ \midrule
Game & ER network &  & 10 min. & 0.1 sec. & Time $\Delta t$ \\ \midrule
Higgs & Followers in social media & Actions on social media & 5 days & 1 sec. & Time $\Delta t$ \\ \midrule
Pems20 & California road network & \multirow{2}{*}{Binarization $\phi$ of average speed} & 2 months & \multirow{2}{*}{5 min.} & \multirow{2}{*}{Time $\Delta t$} \\
Beijing & Beijing road network &  & 6 months &  &  \\ \midrule
Bike & Bike station network & Obs. periods around Covid & 1.5 years & 2 hr. & Bike demand $\phi$ \\ \midrule
\multicolumn{1}{r|}{\multirow{3}{*}{Supply Chain (SC)}} & Production \textit{Plant} & \multirow{3}{*}{Supply chain stages} & \multirow{3}{*}{8 months} & \multirow{3}{*}{1 day} & \multirow{3}{*}{Product amount $\phi$} \\
\multicolumn{1}{r|}{} & Product \textit{Storage} &  &  &  &  \\
\multicolumn{1}{r|}{} & Product \textit{Group} &  &  &  &  \\ \midrule
Baboons* & Face-to-face proximity & Baboon action & 6 weeks & 5 min. & None \\ \bottomrule
\end{tabular}
}
\caption{\textbf{Empirical spreading phenomena on networks across multiple domains}. Description of datasets including network structures, subdatasets induced by features or preprocessing, observation period, minimum temporal resolution, and primary variable used for visualizing results. *The structure of Baboons is temporal and its features have irregular observation periods, instead of controlling for a primary variable we aggregate results over different parameters.}
\label{tab:data}
\end{table}

Our datasets and preprocessing methods yield substantial variability across both structure and data volume. System sizes span more than five orders of magnitude, average degrees tend be lower than 10, while preprocessing choices can shift the amount of data $S$ by several orders of magnitude for the same dataset (\fref{fig:fig4}b). We perform model selection on infection processes by fitting all single mechanisms in our parameter space; however, we found that the independent model was selected in most scenarios (see \ref{sec:si:extrafit}). In empirical data structural information may not capture major aspects of the dynamics (e.g., if the structure is not fully observed, or dynamics depend on exogenous factors or more complex mechanisms). To assess how paradigmatic mechanisms relate to empirical dynamics we also fit noisy versions of the mechanisms that mix an independent component $r$ on the transition probabilities $\hat{F}_{k,m}=r+(1-r)F_{k,m}$. 

We fit both the original and noisy versions of our mechanisms $\mathcal{M}$ and compare loglikelihood ratios pivoting with respect to the most likely model given  $\Delta t$ (\fref{fig:fig4}c). The temporal granularity $\Delta t$ at which we analyze a process can fundamentally reshape model selection, a common trend in our results is for the data to favor different mechanisms depending on the temporal scale. \textit{Game} presents an interesting case study as participants had both individual and group-level incentives (and thus combinations of potential mechanisms), and as a small ER network ($N=36$) it falls within a regime where false positives are likely. Starting from noisy independent infections, decreasing the temporal granularity results in an increasing number of potential mechanisms, starting by threshold models. Such decreasing likelihood ratios, however, capture the combined effect of dynamical ranges and increasingly small amounts of data. In contrast, \textit{Higgs: retweet} is a large network with consistently large amounts of data. While noisy simple infections are generally the favored model, at around $\Delta t\approx 1$ day independent infections dominate. Examining the maximum likelihood parameters displays a qualitative shift in the models as the usually small noise factor becomes more relevant (see \srefsi{sec:si:extrafit}). For the \textit{Pems20} datasets of traffic flow the temporal granularity affects depending on the definition of infected states via the binarization parameter $\phi$. When infected states correspond to low traffic we identify a shift between noisy simple and noisy threshold mechanisms, a behavior that disappears for higher traffic volumes. 

We aggregate over preprocessing parameters and showcase the relative amount of times a particular model was selected, highlighting the vast landscape of potential spreading mechanisms in empirical phenomena (\fref{fig:fig4}d). Preprocessing choices represent modeling assumptions as they can re-frame the processes themselves --- both by determining temporal scales and the definition of binary states. 
While some datasets clearly favor certain families of models, such as noisy simple infections in all \textit{Higgs} subdatasets, arguably similar processes can be explained by vastly different functional forms. For \textit{Baboons}  Affiliative behavior (a combination of pro-social actions) is best explained by complex contagion mechanisms, whereas Attacking actions are spontaneous. Our results are subject to our family of independent mechanisms $\mathcal{M
}$ that don't include common variants (e.g., an Ising process that jointly determines infections and recoveries). However, within our framework simple and threshold models dominate. We find these results striking not only because these mechanisms frame the simple-complex contagion paradigm, but also because both can be present in the same dataset when analyzed at different temporal scales or when defining process differently via binarization, pointing how such aspects are major considerations in understanding empirical phenomena.

\section*{Discussion}
\label{sec:discussion}

Model selection of spreading phenomena on networks deals with a non-trivial interaction of network structure, infection mechanisms and  observation periods. Here, we explore a principled yet flexible framework for the statistical inference of spreading dynamics that includes paradigmatic interaction mechanisms widely explored in the literature (including epidemic dynamics, social contagion, ferromagnetic interactions, and others). We characterize statistical inference in terms of data quantity and the dynamical range of the studied process, i.e. the way data is distributed over structural classes with varying transition probabilities. To instrumentalize model selection, we introduce a measure of model detectability over the dynamical range that is robust to system size, and show that its asymptotic behavior in the thermodynamic limit identifies parameter regions where model selection is more likely to succeed or fail, even for finite systems. Such focus on the dynamical range pinpoints regions of parameter space where model selection might fail despite large system size (and consequently large amounts of observations). Although detectability varies broadly over parameter space, we identify a generic increase with network sparsity (relative number of connections in the system) and closeness to phase transitions in the studied dynamics. We also perform model selection on a wide array of datasets of spreading phenomena on networks, highlighting how the pre-processing of spreading data and its coarse-graining over temporal scales impacts model selection. 

Our work opens up intriguing avenues of theoretical research on the way asymptotic approximations might aid in statistical analysis, as well as on the potential prevalence of idealized spreading mechanisms in real-world complex systems dynamics.
Models of spreading phenomena have a rich tradition in epidemiology~\cite{nowzari_analysis_2016,kiss_mathematics_2017}, contagion and behavior adoption~\cite{granovetter_threshold_1978,yasseri_opinion_2025}, biochemistry~\cite{honey_predicting_2009,haken_chemical_1984}, among many others. Many of these models are simple enough to be interpretable, yet complex enough to capture rich, nonlinear macroscopic behavior~\cite{wigner_unreasonable_1960,watts_simple_2002}. Despite their wide use, to our knowledge, our work represents the first systematic effort to assess both detectability limits and the prevalence of paradigmatic mechanisms in networked spreading phenomena. Our work demonstrates how focusing on dynamical ranges,
we are able to identify scenarios and parameter regimes where conceptually different models may have similar statistical behavior. When using spreading mechanisms to describe, emulate, and predict empirical data, we are thus navigating landscapes of mechanisms that may have similar explanatory power depending on the dynamical regime. Coarse-graining and preprocessing decisions during data handling are not only relevant, but also contextualize regimes where a single mechanism may be neither the best nor the only possible alternative for mathematical description. Understanding this landscape of mechanisms enriches our interpretation of what underlies the phenomena observed for empirical systems.

\section*{Materials and Methods}

{\small \subsection*{Approximation of detectability using asymptotic statistics}
\label{sec:mm:asymdetect}

AMEs are high-accuracy approximations that track the evolution of the relative rates of susceptible $s_{k,m}(t)$ and infected $i_{k,m}(t)$ nodes per $(k,m)$ class. AMEs form a closed system of deterministic equations per degree, 
\begin{equation}
    \begin{split}
        \frac{d}{dt}s_{k,m} &= -F_{k,m}s_{k,m} + R_{k,m}i_{k,m} + \beta_s, \\  
        \frac{d}{dt}i_{k,m} &= -R_{k,m}i_{k,m} + F_{k,m}s_{k,m} + \beta_i,
    \end{split}
\label{eq:ames}
\end{equation}
where $\beta_s$ and $\beta_i$ are nonlinear functions of $(s_{k,m},i_{k,m})$, explored in full elsewhere~\cite{gleeson_high-accuracy_2011,gleeson_binary-state_2013,porter_dynamical_2016,ruan_kinetics_2015,starnini_opinion_2025}. Rates are normalized over degrees, so that the relative fraction of infected nodes of degree $k$ is $\rho_k(t)=\sum_{m=0}^ki_{k,m}(t) = 1 - \sum_{m=0}^ks_{k,m}(t)$, while the overall fraction of infected corresponds to the sum over all degree classes $\rho(t) = \sum_k P_k\rho_k(t)$ where $P_k$ is the degree distribution. The leading terms of Eqs. \ref{eq:ames} correspond to the expected new infections $y_{k,m}=F_{k,m}s_{k,m}$ and recoveries $h_{k,m}=R_{k,m}i_{k,m}$, while $\beta_s$ and $\beta_i$ capture transitions between $m$ classes. 

In contrast with our finite and discrete-time inferential framework, AMEs are approximations in the infinite-size limit and for continuous-time. However, they capture major trends in expectation. For a network of size $N$ the expected number of nodes of degree $k$ is $N_k = N*P_k$; consequently the number of susceptible nodes can be approximated as $S_{k,m}\approx N_ks_{k,m}$. 
Under the AME framework, we estimate the relative ratio of trials over $(k,m)$ classes by numerically calculating the cumulative ratios $\bar{\mathbf{s}}(t^*)=(\bar{s}_{k,m}(t^*))_{k,m}$ where $t^*$ denotes the observation period $[0,t^*]$ and each $\bar{s}_{k,m}(t^*)= \int_{t^*}s_{k,m}(t)dt$ is the cumulative relative ratio of the $(k,m)$ class over $t^*$. The key challenge is that $S_{k,m}$ naturally depends on the network size. For our goal of estimating the dynamic range of a process, i.e., the distribution of susceptible and infected over the $(k,m)$ classes, we use the approximation $S_{k,m} \approx N_k\bar{s}_{k,m}(t^*)$. Larger networks directly impact the expected likelihood ratio (Eq. \ref{eq:elr}) as $S$ is positively associated with system size $N$. The normalizing factor of the $\lambda$ accounts for this dependence $S=\sum_{k,m}S_{k,m} \approx \bar{s}= N \sum_{k,m}P_k\bar{s}_{k,m}$ and yields 
\begin{equation}
    \lambda(\mathbf F|\mathbf{d}_{\omega},\mathbf F_a) = \sum_{k,m}\frac{P_k\bar{s}_{k,m}}{\sum_{k',m'}P_{k'} \bar{s}_{k',m'}}D_{KL}(F_{k,m}||F^a_{k,m}).
\end{equation}

We numerically obtain the cumulative ratios $\bar{\mathbf{s}}(t^*)$ and the cumulative new infections $\bar{\mathbf{y}}(t^*)$ using the $(k,m)$-class expectations, $\bar{y}_{k,m}(t^*)=\int_{t^*}F_{k,m}s_{k,m}(t)dt$. We retain the core shape of the likelihood function induced by the original probability distribution of Eq. \ref{eq:prob}, substituting the dynamic's statistics $(Y_{k,m},S_{k,m})$ for their relative counterparts $(P_k\bar{y}_{k,m}, P_k\bar{s}_{k,m})$. For candidate model $\mathbf F_a$ with transition rates $F^a_{k,m}$, we obtain the parameter's $\theta_{F^a}$ asymptotic maximum likelihood estimate (AMLE), $\hat{\theta}_{F^a} = \max_{\theta_{F^a}}\{l_{AME}(\theta_{F^a}|\mathbf d_{\omega})\}$. Omitting the dependence on $t^*$ for brevity,
\begin{equation}
      l_{AME}(\theta_{F^a}|\mathbf d_{\omega})= - \sum_k P_k \sum_m \bar{y}_{k,m}\log\left(F^a_{k,m}\right) + (\bar{s}_{k,m}-\bar{y}_{k,m})\log\left(1-F^a_{k,m}\right).
\end{equation}
Using numerical optimization tools on Matlab we calculate the AMLEs for the true and candidate model parameters. This method consistently recovers the ground truth parameter $\hat{\theta}_F \approx \theta_F$ across models $\mathbf F$ and structures $\alpha$ (see \ref{sec:si:comparison_synthame}). We obtain the AMLE parameter for all candidate models $\mathbf F_a$, and define the overall detectability as the minimum of all asymptotic likelihood ratios, $\lambda^*(\mathbf F|\mathbf{d}_{\omega})=\min_{\mathbf F_a}\{\lambda(\mathbf F|\mathbf{d}_{\omega}, \mathbf F_a)\}$. The candidate set typically includes the seven alternative mechanisms, though we exclude Ising models from each other's candidate sets to prevent reporting them as undetectable --- an artifact of separating low- and high-temperature regimes.

}

{\small \subsection*{Heuristics for identifying transient/stationary dynamics}
\label{sec:mm:paramspace}

We use heuristics to identify transience and stationarity in both simulations on synthetic networks and asymptotic approximations.
For synthetic networks we define transience as twice the relaxation time $\tau_e$, an  approximation of decay for the autocorrelation function $C(\tau)$. Let $\{I(t)\}$ be the sequence of infected counts over times $t=0, \ldots, T_{\mathrm{max}}$ that result from a random realization of $\omega \in \Omega$ on a network of size $N$. The normalized autocorrelation function measures correlation between infected values at a lag $\tau$,  
\begin{equation}
C(\tau) = \frac{\langle I(t)I(t+\tau)\rangle - \langle I \rangle^2}{\langle I ^2\rangle- \langle I \rangle^2}.
\end{equation}
We identify the relaxation time $\tau_e$ (and a coefficient $c$) assuming exponential decay by fitting the observed autocorrelation $C(\tau)$ to the functional form $C(\tau)\approx ce^{-\tau/{\tau_e}}$. For our purpose, transience refers to observation periods where the dynamical range broadly changes; by fitting exponential decay we do not explicitly consider scenarios such as critical behavior where a different functional form may be a better fit. However, we find that consistently defining transience under exponential decay maintains dynamical ranges consistent over system sizes (\fref{fig:fig2}c). We analyze additional definitions of transience in SI.

We implement our asymptotic analysis on Matlab, where we partition $\Omega$ by linearly dividing the pre-determined parameter ranges ($\theta_F, \theta_R, i_0$) among 15 points, and evaluating 20 values $\langle k \rangle$ (see ranges in SI). For each point $\omega \in \Omega$ we track the evolution of $\mathbf{s}(t)=(s_{k,m}(t))$ and $\mathbf{i}(t)=(i_{k,m}(t))$ for a fixed evolution period $T=100$. We estimate the transient period $t^*$ by identifying first value $t'$ such that infectivity rates remain bounded for the rest of the observation period $t^* = \{t': |\rho(t')-\rho(t)|<\epsilon,  \forall t\geq t'\}$. We fix the bound at $\epsilon=10^{-4}$. As discrete and finite-time systems differ from continuous, infinite-size approximations, we find our two heuristics to work well in practice. Fitting exponential decay aids in minimizing fluctuations of detectability over the five orders of magnitude we analyze, $N=10, \ldots, 10^5$. Yet not assuming exponential decay on our asymptotic analysis allows for assessing the effect near phase transitions.   
Additional implementation details are available at \ref{sec:si:ames}. 
}

{\small \subsection*{Empirical spreading data}

We analyzed datasets of empirical spreading phenomena where both a network structure and timestamped dynamics were available. \textit{Game} is a game-theoretic experiment in which 36 participants interacted on a static Erdős–Rényi network, aiming to reach consensus in a binary-state game~\cite{kearns_behavioral_2009}. Each participant could observe the current states of their neighbors but also had individual incentives favoring one of the two states. \textit{Higgs} tracks information spread regarding the Higgs boson discovery on an open social media platform~\cite{de_domenico_anatomy_2013}. The structure corresponds to follower links, and the dynamics to activity on the topic. \textit{Urban} contains traffic flow on the road networks of two cities (\textit{Pems20} and \textit{Beijing}), as well as \textit{Bike}, a public-bicycle demand where the structure is based on station proximity~\cite{cai_urbandata_2023}. \textit{SupplyChain} are the daily aggregated supply-chain interactions within a large factory, where nodes are products and edges reflect co-production, co-location, or grouping~\cite{wasi_supplygraph_2024}. Node features are different measures of product quantity: production, delivery, factory issues or client orders. \textit{Baboons} captures the behavioral dynamics of 16 primates, recorded via Bluetooth proximity sensors to form a contact network~\cite{gelardi_detecting_2019}. Their behavior was classified by researchers into discrete behavioral categories at particular times during the day  during a month. For all datasets we homogenize preprocessing pipelines so that datasets can be categorized in terms of sub-datasets that can be analyzed by varying a primary preprocessing parameter --- one that controls the temporal granularity of observations, and another that binarizes continuous state features. The algorithms used for binarization can be found in SI.

}

{\small \subsubsection*{Data and code availability}
Code to reproduce the results of the paper is available. 
Data is openly available from their original sources (preprocessed \textit{Game} available in our repository).

\subsubsection*{Acknowledgments}
J.U.C. and G.I. thank Giulia de Meijere for valuable suggestions. 

\subsubsection*{Author contributions}
J.U.C., T.P., and G.I. conceived, designed, and developed the study.  J.U.C. derived analytical results, performed numerical simulations and model fitting, and analyzed empirical data. J.U.C., T.P., and G.I. wrote the paper.

\subsubsection*{Competing interest statement}
All authors declare no competing interest.

\printbibliography

\clearpage
\appendix
\begin{center}
{\LARGE Supplementary Information for}\\[0.7cm]
{\Large \textbf{Statistical inference of dynamical processes on networks}}\\[0.5cm]

{\large J. Ureña-Carrión$^*$, T. P. Peixoto, G. Iñiguez$^*$}\\[0.7cm]
{\small $^*$Corresponding author email: javier.urenacarrion@tuni.fi, gerardo.iniguez@tuni.fi}\\[1cm]
\end{center}

\addtocontents{toc}{\protect\setstretch{0.1}}
\tableofcontents

\section{Derivation of expected log-likelihood ratios}
\label{sec:si:theo_deriv}

Within our framework, let $\mathbf{D}=(\mathbf{Y}, \mathbf{S}, \mathbf{H}, \mathbf{I})$ denote data and  $(\mathbf{F}, \mathbf{R})$ be a binary-state model with Let $\mathbf{F}=\{F_{k,m}(\theta_F)\}_{k,m}$ and $\mathbf{R}=\{R_{k,m}(\theta_R)\}_{k,m}$  parameter $\theta=(\theta_F,\theta_R)$ and where  $Y_{k,m}\sim\mathrm{Binomial}(S_{k,m}, F_{k,m})$ and $H_{k,m}\sim\mathrm{Binomial}(I_{k,m}, R_{k,m})$. The loglikelihood function is given by


\begin{equation}
    \begin{split}
        l(\theta|\mathbf{D}) &= \sum_k \sum_m Y_{k,m}\log(F_{k,m}(\theta_F)) + (S_{k,m}-Y_{k,m})\log(1-F_{k,m}(\theta_F)) +\\
         & H_{k,m}\log(R_{k,m}(\theta_R)) + (I_{k,m}-H_{k,m})\log(1-R_{k,m}(\theta_R)). 
    \end{split}
\end{equation}





The likelihood function can be used for any family of models of binary trials that can be expressed in terms of transition rates $F_{k,m}$ and $R_{k,m}$. In our framework, all spreading mechanisms depend only on one parameter, which allows for direct comparisons between models using the likelihood ratio. Let 
$\mathbf{F}^a=\{F^a_{k,m}(\theta_{F^a})\}_{k,m}$ denote the rates of an alternative mechanism. 
Assuming that the recovery mechanism is fixed, i.e., that we're focusing on model selection on the infection mechanism only, the loglikelihood ratio of these two models can be written as the difference between their loglikelihoods $L_R(\theta_F, \theta_{F^a}|\mathbf{D_F})=l(\theta_F|\mathbf{D_F})-l(\theta_{F^a}|\mathbf{D_F})$. This expression can be restated using the ratios of transition rates per $(k,m)$ class. For simplicity, here $F_{k,m}=F_{k,m}(\theta_F)$ and $F^a_{k,m}=F^a_{k,m}(\theta^a_F)$, 
\begin{equation}
    \begin{split}
        L_R(\theta, \theta^a|\mathbf{D}_F) &= \sum_{k,m} Y_{k,m}\log\left(F_{k,m}\right) + (S_{k,m}-Y_{k,m})\log\left(1-F_{k,m}\right) - \\
        &\left(\sum_{k,m} Y_{k,m}\log\left(F^a_{k,m}\right) + (S_{k,m}-Y_{k,m})\log\left(1-F^a_{k,m}\right) \right)\\
        &= \sum_{k,m} Y_{k,m}\log\left(\frac{F_{k,m}}{F^a_{k,m}}\right) + (S_{k,m}-Y_{k,m})\log\left(\frac{1-F_{k,m}}{1-F^a_{k,m}}\right) \\
    \end{split}    
\end{equation}

Noting that for our binomial trials $\mathbb{E}[Y_{k,m}]=F_{k,m}S_{k,m}$, applying the operator on new infections $Y_{k,m}$ we get expected log-likelihood ratio is thus,

\begin{equation}
    \label{eq:exp_llikr}
    \begin{split}
        \mathbb{E}[L_R(\theta_F, \theta_{F^a}|\mathbf{D}_F)] &= \sum_{k,m} \mathbb{E}[Y_{k,m}]\log\left(\frac{F_{k,m}}{F^a_{k,m}}\right) + \mathbb{E}[S_{k,m}-Y_{k,m}] \log\left(\frac{1-F_{k,m}}{1-F^a_{k,m}}\right) \\
        &= \sum_{k,m} S_{k,m}F_{k,m}\log\left(\frac{F_{k,m}}{F^a_{k,m}}\right) + S_{k,m}(1-F_{k,m}) \log\left(\frac{1-F_{k,m}}{1-F^a_{k,m}}\right) \\
        &=\sum_{k,m}S_{k,m}D_{KL}(F_{k,m}||F^a_{k,m})
    \end{split}
\end{equation}

Where $D_{KL}(P||Q) = \sum_xP(x)\log\left(\frac{P(x)}{Q(x)}\right)$ is the Kullback-Leibler (KL) Divergence. This function is not defined when $Q(x)=0$ and $P(x)>0$, or $Q(x)=1$ and $P(x)<1$. Detectability is the expected loglikelihood ratio normalized using the total amount of data $S = \sum_{k,m}S_{k,m}$; we use notation of models $\mathbf{F}$ instead of parameters $\theta_F$ to highlight its role in model selection, 

\begin{equation}
    \Lambda(\mathbf{F}, \mathbf{F}^a|\mathbf{D}_F) = \frac{\mathbb{E}[L_R(\theta_F, \theta_{F^a}|\mathbf{D}_F)]}{S}.
\end{equation}

\section{Experimental setup of parameter space}
\label{sec:si:mechanisms}

We select a set of six spreading mechanisms that are commonly used in the literature: Independent, Simple, Voter, Majority Voter, Threshold and Ising, including both Absolute and Relative threshold models and high- and low-temperature Ising models; $T=10$ and $T=1$, respectively. Our framework is symmetric with respect to the infected class, meaning that $X=0$ and $X=1$ are competing but interchangeable states. In practice, this means that $R_{k,m}=F_{k,k-m}$ --- a node in state $X=0$ is considered susceptible to infected nodes $X_a=1$, but a node in state $X=1$ is considered "susceptible" to "infected" nodes $X_a=0$. Figure \ref{fig:mechanisms}
exemplifies the transition probabilities as a function of number of infected neighbors $m$ for a node of degree $k=15$. Mechanisms capture largely distinct functional forms, although some parameterizations might lead to similar descriptions. For instance, the Ising, threshold and Majority models are al implemented using sigmoid functions $\phi$ to ensure differentiability. They all parameterize different relationships within the sigmoid --- the threshold $h$ refers to the absolute or relative number of infected neighbors necessary to transition, whereas in the majority voter the threshold is always maintained at a simple majority $m > k/2$, but we parameterize a baseline level of noise $q$. This means that for some parameter values the functional forms of different mechanisms are similar. These scenarios are evident, for example, when looking at the detectability of particular parameter values further explored on Figure \ref{fig:ame-thetak-detect}. 

\begin{figure}
    \centering
    \includegraphics[width=0.85\linewidth]{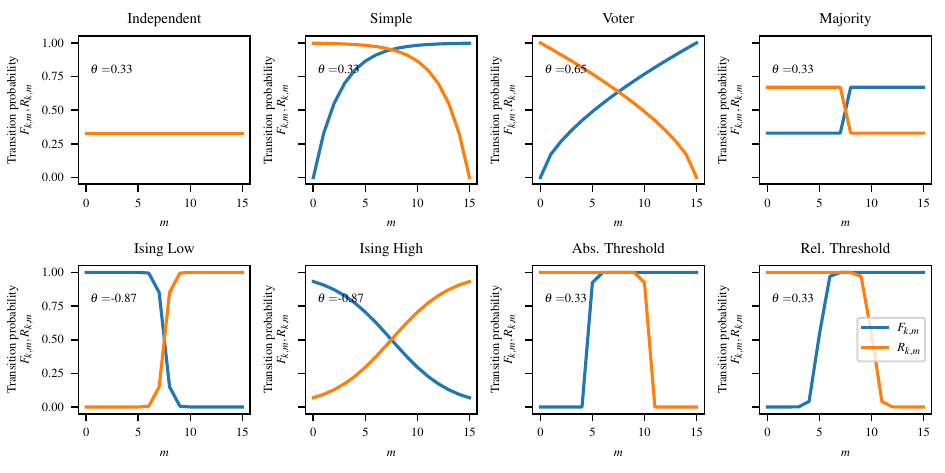}
    \caption{\textbf{Transition probabilities for a node of degree $k=15$.} Each panel depicts the transition probabilities $F_{k,m}$ and $R_{k,m}$ for a node of degree $k=15$ as a function of the number of infected neighbors $m$, for fixed model parameters ($\theta_F$ or $\theta_R$, annotations). }
    \label{fig:mechanisms}
\end{figure}

We avoid issues of indeterminacy by adding small bounds via redefined the transition rates under a simple transformations of the transition rates, $F_{k,m}:=(1-2\epsilon)F_{k,m}+\epsilon$, where $F_{k,m}$ are the original transition rates. We use $\epsilon=10^{-5}$ in all our calculations, which leads to an upper bound on the KL divergence ($D_{KL}\lessapprox 16.6$ in our case). These transformations have a largely negligible impact on our results as we focus on models where divergence is small. However, they act as small baseline noise, and thus allow us to compare comparing functional forms over the same domain. The alternative of not including $\epsilon$ would result in some models being rejected a priori (e.g., models that assign probability zero or one to some events), without considering whether they approximate some other model's dynamical range in an otherwise accurate manner. 

\begin{table}[]
\footnotesize
\centering
\begin{tabular}{c|ccc}
\textbf{Category}           & \textbf{Subcategory} & \textbf{Parameter} & \textbf{Values}             \\ \hline
\multirow{3}{*}{Topology}  
    & ER  & $\langle k\rangle$   & $0.5, 0.75 ,1, 1.5, 2, 3, 5, 8, 10, 13, 17, 22, 28, 32$    \\
    & Log-Normal & $\sigma^2_k$ & 1, 2, 2.5, 3, 3.5, 4, 5, 6, 8, 10, 12, 20, 35, 50    \\ \hline 
Initial Conditions  & / & $i_0$ & $(0, 1)$  \\ \hline
\multirow{6}{*}{Mechanisms} 
    & Independent & $r$ & $(0, 1)$ \\
    & Voter  & $\gamma$ & $[0.1, 1.9]$ \\
    & Majority & $q,\alpha=50$ &$[0, 1]$ \\
    & Simple & $d$ & $(0, 1]$ \\
    & Ising Low & $J,T=1$ & $[-2.25, 2.25]$ \\
    & Ising High & $J,T=10$ & $[-2.25, 2.25]$ \\
    & Abs. Threshold & $h,\alpha=50$ & $(0, 1)$ \\
    & Rel. Threshold & $h,\alpha=50$ & $(0, 1)$ \\\hline
\end{tabular}
\caption{Partition of our parameter space  $\Omega$ and discretization used for the systematic analysis. The recovery rates are symmetric on the number of infected neighbors $R_{k,m}=F_{k,k-m}$, we partition the parameter ranges by using 15 equally-spaced values. Closed ranges $[a,b]$ include $a$ and $b$ on partition of the space; open ranges  $(a,b)$ maintain the same partition but include a buffer of size $10^{-3}$, as $a$ or $b$ are parameterizations where mechanisms are not meaningfully present (e.g., no transitions). }
\label{tab:models_params}
\end{table}

In our parameter space we decouple system size $N$ from degree distribution, which we parameterize $\alpha$ in terms of density using a Poisson distribution for ER networks $\{p_k\}\sim \text{Poisson}(\langle k\rangle)$. For the log-normal we use a discretized lognomal distribution with a fixed average degree $\mu_k=3$. Depending on the implementation (Python, Matlab), fixing the average degree requires identifying the lognormal location and scale parameters $\mu_K, \sigma_K$. 

\subsection{Detectability on synthetic networks}
\subsubsection*{Simulations}
\label{sec:si:synth-detect}

We implemented a pipeline for simulations on synthetic networks that includes data simulation and statistical inference. To generate data we first obtain a realization of a random graph of predetermined size $N$ and degree distribution $\{p_k\}$. We initialize network states by independently setting state $X_i=1$ with probability $i_0$. At each time step, nodes change state, depending on their current state and the infection $\mathbf{F}$ and recovery $\mathbf{F}$ models. At each step, we track data aggregated on topological classes $\mathbf{D}=(\mathbf{Y,S,H,I})$, including current amounts of susceptible $\mathbf{S}$ and infected $\mathbf{I}$, as well as new infections $\mathbf{Y}$ and new recoveries $\mathbf{H}$. 

To assess the role of systems size on detectability and expected loglikelhood ratios, on Figure \ref{fig:detect_components} we plot a sample of 250 points $\omega \in \Omega$ and compute their standard deviation when increasing networks size for all models. Detectability becomes increasingly stable with regards to the detectability of the largest system, and its standard deviation decreases. In contrast, the standard deviation of expected loglikelihood ratios (the unnormalized detectability $S\times\Lambda$) increases for larger systems,  hindering their suitability for being approximated asymptotically. 

\begin{figure}
    \centering
    \includegraphics[width=\linewidth]{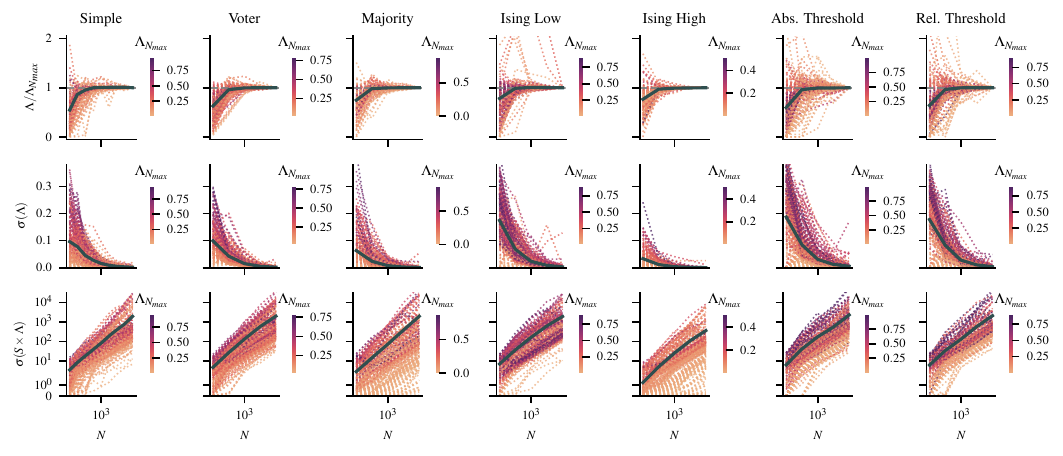}
    \caption{\textbf{Detectability stabilizes with system size, whereas expected likelihood ratios diverge}. We compute the trajectories (dotted lines) of 250 points $\omega \in \Omega|\mathbf{F}$ when increasing system size $N$ for all models (\textit{columns}). Each dotted line is an average of 10 realizations of the same point $\omega$ per system size [total 50 realizations per line], solid line is average of trajectories. (\textit{top}) Detectability relative to the largest systems $\Lambda/\Lambda_{N_{max}}$  tends to stabilize, including (\textit{middle}) a decreasing  standard deviation. In contrast, (\textit{bottom}) expected loglikelihood ratios $S\times \Lambda$ diverge as their standard deviation increases for larger $N$. For all models $\mathbf{F}^a$ is Independent.}
    \label{fig:detect_components}
\end{figure}

\subsubsection{Additional Transience Heuristics}
\label{sec:si:extra-heuristics}

In the main text we identify transient periods by fitting exponential decay on the autocorrelation function. We explore the behaviour of detectability when we define stationarity when either (a) the process has decayed, or (b) the number of new infections and new recoveries are in balance $Y(T^*)\approx H(T^*)$ (for convenience we call relaxation time). We determine $T^*$ by first identifying whether the dynamics have reached an absorbing state (only for models that don't include spontaneous state changes), or otherwise using a hypothesis test with z-scores for difference of means. We define the new infection and recovery rates $\bar{Y}(t)=\frac{1}{N}\sum_{k,m}Y_{k,m}(t)$, and $\bar{H}(t)=\frac{1}{N}\sum_{k,m}H_{k,m}(t)$, respectively. And the total state change rate $V(t)=\frac{Y(t)+H(t)}{2N}$. The z-score at step $t$ is

\begin{equation}
z(t)=\frac{|\bar{Y}(t)-\bar{H}(t)|}{\sqrt{V(t)(1-V(t))(\frac{1}{N}+\frac{1}{N})}}
\end{equation}

We fix the value for determining $T^*$ at $z^*=1.285$ while noting that this value consistently leads to shorter relaxation times for smaller systems. We use a fixed value $z^*$ as it is generally difficult to determine stationarity for small systems where fluctuations are large with respect to the system itself. The question of transience for different systems sizes is relevant as we can expect the dynamic range to vary in transient states, and largely stabilize in stationary dynamics (we note that in this scenario we do not distinguish between, e.g., a stationary state that oscillates and a transient state, as we largely focus on whether the dynamic range is balanced to some extent). Instead of using size-dependent thresholds, which would add a new functional relationship over a large parameter space, we briefly explore how system size and observation periods relate. In Figure \ref{fig:si:detect_station} we present a comparison between a random samples of points aggregated over transient periods (obtained with z-score $z^*=1.285$) and fixed for $T=10$. Our results show that the trend of increasing but stabilitzing detectability holds for both fixed and transient observation periods. Although relaxation times are generally shorter for smaller $N$, by testing for fixed times $T$ the loss of detectability is more likely to be a byproduct of system size, and not observation periods. The largest difference is for smaller systems: they consistently display more heterogeneous detectability using transient periods, but also display smaller average changes with respect to the large-scale detectability. 

\begin{figure}
    \centering
    \includegraphics[width=\linewidth]{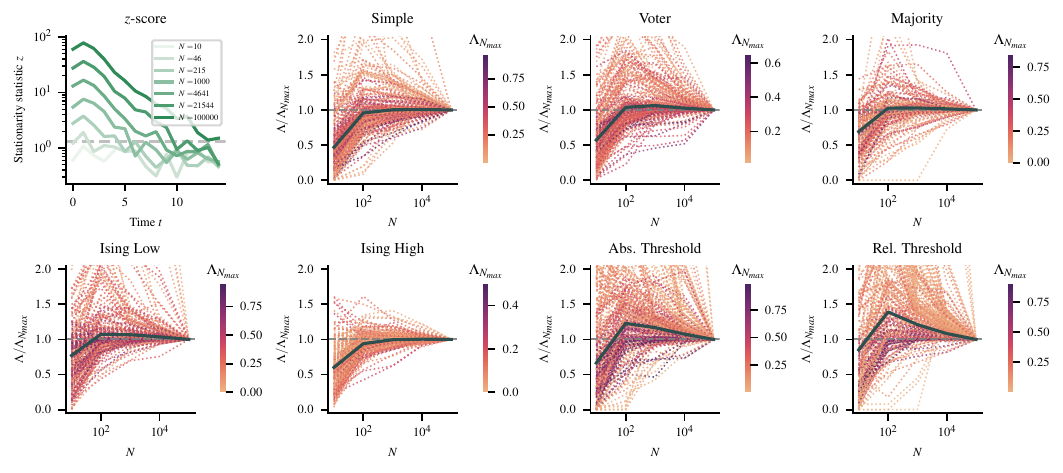}
    \caption{\textbf{Transience as balance between new infections and recoveries}. (\textit{top left}) Stationarity statistic $Z$ as a function of time and system size $N$ for a point in the parameter space that doesn't usually die out ($\mathbf{F}$ Simple, $\mathbf{R}$ Voter, $\langle k \rangle=6$, $\theta_F=.37$, $\theta_R=.9$, $i_0=.1$), dotted line $z=1.285$. Larger systems generally reach a balance between infections and recoveries at a slower rate than their smaller counterparts. (\textit{Subplots}) Relative detectability $\Lambda/\Lambda_{N_{max}}$ for 250 random points $\omega$ for models using $z$-score transient observation periods. Under the $z$-score heuristic the smallest systems are still less detectable, yet mid-size systems might display higher detectability values, and trajectories display higher variability generally. }
    \label{fig:si:detect_station}
\end{figure}

\subsubsection*{Inference}

Given data $\mathbf{D}$ we obtain Maximum Likelihood (ML) estimators with automatic differentiation on Python to identify the optimal parameter $\theta=(\theta_F, \theta_R)$ for model $(\mathbf{F},\mathbf{R})$, so that, $\hat{\theta}=\arg \max_{\theta} \{l(\theta|\mathbf{D})\}$. Our results focus on inference of infection mechanisms that are independent from the recoveries, so in most of our results we use only infection data $\mathbf{D}_F=(\mathbf{Y}_F, \mathbf{S}_F)$. 
When including results for  \textit{detectability} where the  true generative mechanism $(\mathbf{F}, \theta_F)$ we minimize over the alternative model's parameter $\theta_C$, i.e., $\Lambda=\min_{\theta_C}\{\sum_{k,m}S_{k,m}D_{KL}(F_{k,m}(\theta_F)||F^C_{k,m}(\theta_C))\}$. 
However, when performing model selection we perform likelihood ratio tests by first calculating optimal ML parameters for all mechanisms $\mathcal{M}$ (including for the true generative mechanism) and use the estimated ML parameters $\hat{\theta}_F$ and $\hat{\theta}_C$ on the loglikelihood functions, $LR(\mathbf{F}, M_C)=l(\hat{\theta}_F|\mathbf{D_F})-l(\hat{\theta}_C|\mathbf{D_F})$, and reject the models if they are the test is positive. Section \ref{sec:si:comparison_synthame} includes results the performance of our approach when inferring a true parameter $\theta_F$ using ML estimators $\hat{\theta}_F$. 

\subsection{Implementation of AME framework}
\label{sec:si:ames}

Approximate master equations (AMEs) focus on the probabilities that a node of degree $k$ is infected or recovered, respectively, when $m$ of its neighbors are infected \cite{vazquez_systems_2008,ruan_kinetics_2015,gleeson_high-accuracy_2011,gleeson_binary-state_2013}. This approximates well the full opinion dynamics for large configuration-model networks \cite{newman_networks_2018}. We write $s_{k, m}(t)$ for the fraction of nodes with degree $k$ that are susceptible at time $t$ (i.e. holding opinion $s = 0$) and have $m$ infected neighbors. Similarly, $i_{k,m}(t)$ is the proportion of degree-$k$ nodes that are infected and have $m$ infected neighbors. Thus we have $\sum_{m=0}^k [s_{k,m}(t)+i_{k,m}(t)]=1$ for all $k$. The fraction of infected nodes among all degree-$k$ nodes is $\rho_k(t) = \sum_{m=0}^k i_{k, m} = 1 - \sum_{m=0}^k s_{k, m}$, while the fraction of infected nodes in the entire network is $\rho(t) = \sum_{k=0}^\infty P_k \rho_k$, with $P_k$ the degree distribution of the network. 

The AMEs are coupled differential equations for $s_{k, m}$ and $i_{k, m}$. The rates at which a node of class $(k,m)$ changes from susceptible to infected or vice versa are given by $F_{k,m}$ and $R_{k,m}$. A node may also change, say, from $(k,m)$ to $(k,m+1)$ if one of its neighbors becomes infected. To formulate a closed set of equations for the $s_{k,m}$ and $i_{k,m}$, \textcite{gleeson_high-accuracy_2011,gleeson_binary-state_2013} proceeds by assuming there are global rates with which links of type $SS$ change to $SI$, $II$ to $SI$, or $SI$ to either $SS$ or $II$ respectively (where $S$ and $I$ denote the opinions of nodes at the endpoints of an edge). These rates can be expressed in terms of $s_{k,m}$ and $i_{k,m}$, by using $F_{k,m}$ and $R_{k,m}$ and disregarding higher-order correlations. The end result is a closed set of differential equations for $s_{k,m}$ and $i_{k,m}$. In a network of maximum degree $k_{\rm max}$, the AME system has $(k_{\rm max} + 1) (k_{\rm max} + 2)$ degrees of freedom \cite{ruan_kinetics_2015,porter_dynamical_2016}. 

\subsubsection*{Asymptotic detectability pipeline}

We use AMEs to obtain the asymptotic loglikelihood ratio over alternative models. As with synthetic networks, our approach is to first simulate data $\mathbf{d}_{\omega}=(\mathbf{y}_{\omega},\mathbf{s}_{\omega},\mathbf{h}_{\omega},\mathbf{i}_{\omega})$ for a $\omega\in \Omega$, aggregated over transient and stationary periods. AMEs approximate expected values over topological classes, so that both new infections $y_{k,m}(t)=s_{k,m}(t)F_{k,m}$ and new recoveries $h_{k,m}(t)=i_{k,m}(t)R_{k,m}$ are stated in terms of expected values. In Algorithm \ref{alg:ame_pipeline} we describe the major steps of our pipeline for determining the asymptotic detectability of a point in our parameter space $\omega\in \Omega$. We implemented this  pipeline on MATLAB, adapting the AME solvers from \cite{gleeson_binary-state_2013}. In this section we further describe the major steps involved in the process. 

\begin{algorithm}
\caption{Pipeline for AME framework}
\label{alg:ame_pipeline}
\begin{algorithmic}[1]
\Require Point in parameter space $\omega \in (\alpha, \mathbf{F}, \mathbf{F}, \theta_F, \theta_R, i_0)$, maximum observation period $t_{max}$, transience tolerance $\epsilon$.
\State Obtain degree distribution $\{p_k\}_k$ induced by $\alpha$. 
\State Numerically solve AMEs induced by $\omega$ for time range $T=[0, t_{\rm max}]$, obtain tensors over $(k,m)$ classes and time range for fractions of susceptibles $[\mathbf{s}(t)]_T$, new infections $[\mathbf{y}(t)]_T$, infections $[\mathbf{i}(t)]_T$ and new recoveries $[\mathbf{h}(t)]_T$.
\State Identify relaxation time $t^* = \{t': |\rho(t')-\rho(t)|<\epsilon,  \forall t\geq t'\}$.
\State Obtain cumulative distributions $\bar{\mathbf{d}}_{\omega}(t^*)$, including  susceptibles $\bar{\mathbf{s}}$ and new infections $\bar{\mathbf{y}}$.
\State AML estimate $\hat{\theta}_F$ for true infection mechanism using asymptotic data $\bar{\mathbf{d}}_{\omega}$.
\For{alternative mechanism $\mathbf{F}^a \in \mathcal{M}_{}\backslash\{\mathbf{F}\}$}
    \State AML estimate for alternative mechanisms $\hat{\theta}_{F^a}$. 
    \State Asymptotic detectability under alternative mechanism $\lambda(\mathbf{F}|\mathbf{d}_{\omega},\mathbf{F}^a(\hat{\theta}_{F^a}))$.
    \EndFor
\State Identify minimum detectability $\lambda^*(\mathbf{F}|\mathbf{d}_\omega)$.
\State \Return Process statistics including final state $\rho$, relaxation time $t^*$, amount of data $(\bar{s},\bar{y})$, AML parameter estimates $\{\hat{\theta}\}_{M}$ and asymptotic detectabilities $\{\lambda\}_{\mathbf{F}^a}$.  
\end{algorithmic}
\end{algorithm}

We identify the range of degrees as those values that capture at least $0.9995$ cumulative probability, i.e., $\sum_{k_{\rm min}}^{k_{\rm max}}p_k \geq 0.9995$. For ER networks the sum is centered around the average value $\langle k \rangle$, while for LR networks the lower value is fixed $k_{\rm min}=1$. We simulate each point $\omega$ of our partitioned parameter space over a period $t_{\rm max}=100$ and calculate the cumulative relative susceptible and infected nodes over the relaxation time $t^*$ as the integral $\bar{s}_{k,m}(t^*)=\int_{t=0}^{t^*}s_{k,m}(t)dt$. For AMEs the relaxation time is defined as the first moment $t^*$ where $\frac{ds_{k,m}(t^*)}{dt}=0$ and $\frac{di_{k,m}(t^*)}{dt}=0$ for all $(k,m)$ classes. 


\subsection{Inference using  simulated data and asymptotic approximations}
\label{sec:si:comparison_synthame}

We approximate infection data $\mathbf{D}_\omega=(\mathbf{Y}_\omega,\mathbf{S}_\omega)$ aggregated over a transient period of $\omega \in \Omega$ using numerical integration under AMEs, $\mathbf{d}_{\omega}(t^*)=(\mathbf{y}_{\omega}(t^*), \mathbf{s}_{\omega}(t^*))$. Both $T^*$ and $t^*$ are approximated using heuristics of stationarity.

\begin{figure}
    \centering
    \includegraphics[width=\linewidth]{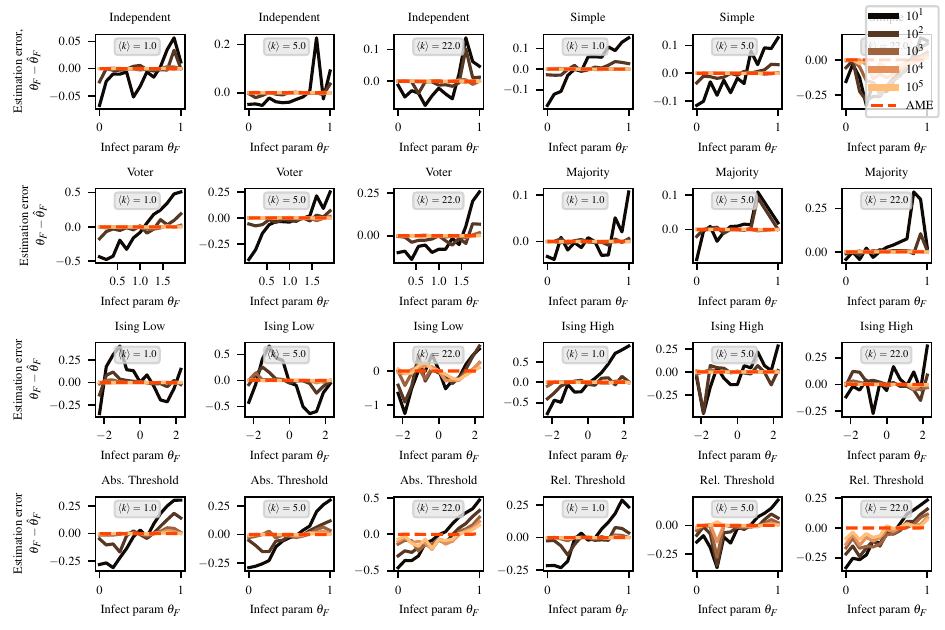}
    \caption{\textbf{Parameter estimation error using simulations and AMEs}. Each subplot depicts the median estimation error for the infection parameter $\theta_F$ comparing (colors) different networks sizes and data from AMEs. Each mechanism $\Omega$ has three subplots corresponding to topologies $\langle k \rangle=1,5,22$. Errors sometimes follow trends dependent on $\theta_F$, particularly for small systems $N=10,100$, yet errors are largely negligible for larger systems (overlapped around zero), with some exceptions for dense networks (e.g. threshold models). Estimations using AMEs (red dotted line) show a mean estimation error close to zero over all parameter ranges. }
    \label{fig:si:param-space}
\end{figure}

The total number of susceptible nodes $S_{k,m}$ depends on the observation period. This means that estimated relaxation times using AMEs $T^{AME}$ and discrete simulated data $T^S$ affect the detectability of their respective processes. A key question for the validity of our process is thus to assess the extent to which AMEs can be used to estimate detectability under these conditions. We use two different methods to estimate these two quantities per point of the parameter space $T^{AME}$ and $T^S$.

We approximate the total number of infected per $(k,m)$-class. 
One reason why we normalize with $S=\sum_{k,m}S_{k,m}$ is to focus on how the mechanisms differ (their weighed divergences), as opposed to the amounts of susceptible (how large the $S_{k,m}$ are). However, we can also use AMEs to approximate a certain network size, so that $p_k=\frac{N_k}{N}$. In this case, we approximate the number of infected as $S_{k,m}\approx N_ks^A_{k,m}=Np_ks^A_{k,m}$. 

\begin{figure}
    \centering
    \includegraphics[width=\linewidth]{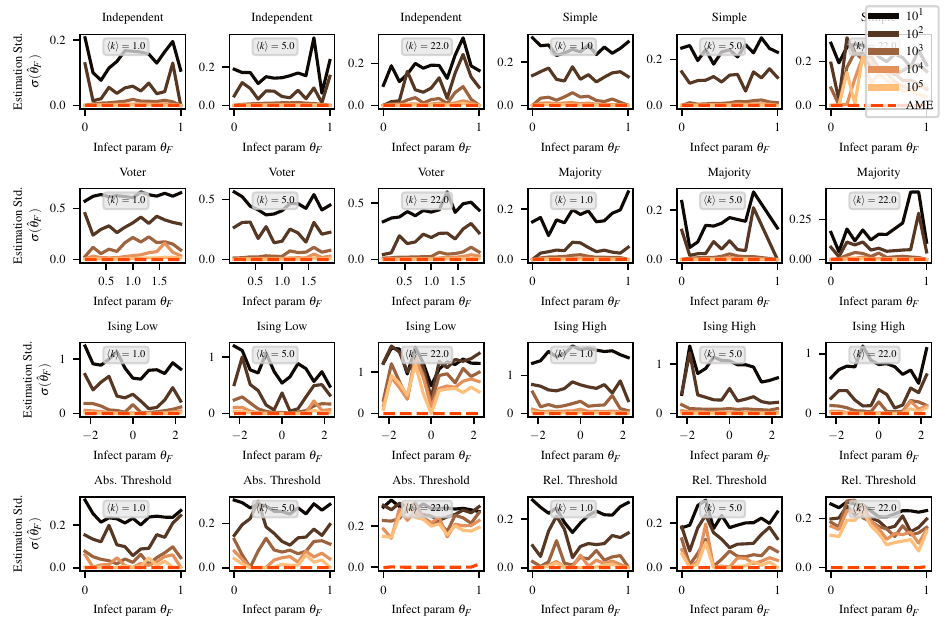}
    \caption{\textbf{Standard deviation of infection parameter estimate $\hat{\theta_F}$ over different runs and points the $\Omega|\theta_F, \mathbf{F}$}. Each subplot depicts the median estimation error for the infection parameter $\theta_F$ comparing (colors) different networks sizes and data from AMEs. Each mechanism $\Omega$ has three subplots corresponding to topologies $\langle k \rangle=1,5,22$. Standard deviations largely follow similar trends with larger values for smaller systems and large systems closer to zero. However, for Ising Low and the threshold models even large systems display larger standard deviations sometimes follow trends dependent on $\theta_F$, particularly for small systems $N=10$. For Majority, the errors are pronounced in all systems sizes and lower density networks. Estimations using AMEs (red dotted line) show a mean estimation error close to zero over all parameter ranges. }
    \label{fig:si:param-space}
\end{figure}

\section{Additional results for asymptotic detectability}

\subsection{All models and degree heterogeneity}
\label{sec:si:ames_add}

AMEs are a high-accuracy approximations for dynamics on networks. In the binary case, they track the relative fraction of susceptible $s_{k,m}(t)$ and infected nodes $i_{k,m}(t)$ for each $(k,m)$-class. This relative fraction is per degree $k$, so that AMEs meet the constraint $\sum_{m=0}^ks_{k,m} + \sum_{m=0}^ki_{k,m}= 1$ per degree  $k$. 

\begin{figure}
    \centering
    \includegraphics[width=\linewidth]{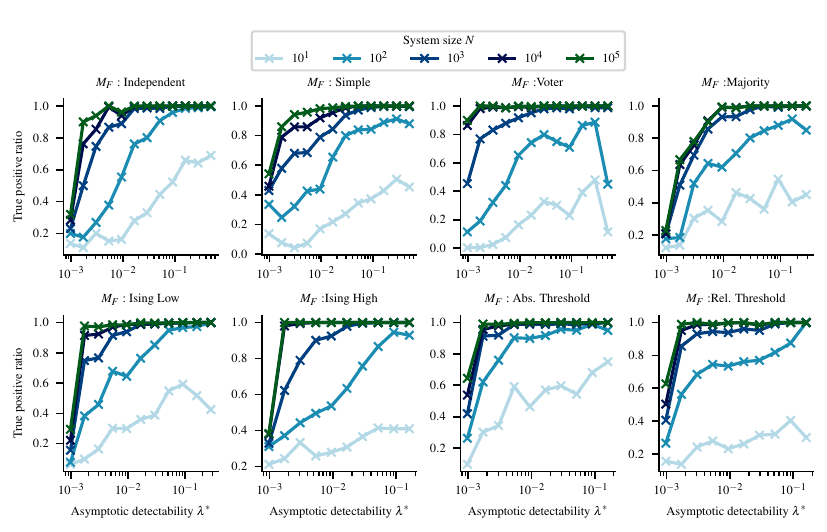}
    \caption{\textbf{Asymptotic detectability and model selection for synthetic data}. For eight spreading mechanisms $\mathbf{F}$ (each subplot), we systematically analyze a partition of the parameter space $\Omega|\mathbf{F}$ and obtain asymptotic detectability values $\lambda^*$. We obtain of points of a given asymptotic detectability. For each point, we simulate processes where we range system sizes over five orders of magnitude $N=10^1, \ldots, 10^5$, and obtain the True Positive Ratio on synthetic networks. For most mechanisms our results show  higher TPRs for higher asymptotic detectability and system size. These results, however, do not necessarily hold for higher $\lambda^*$ in the Voter model and the Majority Voter model. }
    \label{fig:si:param-space}
\end{figure}

We use AMEs to estimate the detectability of spreading mechanisms over a wide parameter space. We use the expected log-likelihood ratio $\mathbb{E}[LR(\theta, \theta^a)]$ (Eq. \ref{eq:exp_llikr}) as a measure of detectability. This measure has two main components: the way in which mechanisms differ per $(k,m)$ class (the divergence component $D_{KL}(F_{k,m}||F_{k,m}^a)$), and the number of exposed nodes in that class ($S_{k,m}$ or $I_{k,m}$). One possible interpretation is that even processes that are very similar (a small but positive divergence for some $(k,m)$ classes) can be separated if there are enough observations (the corresponding $S_{k,m}$ values are large enough).


\begin{figure}
    \centering    
    \includegraphics[width=0.85\linewidth]{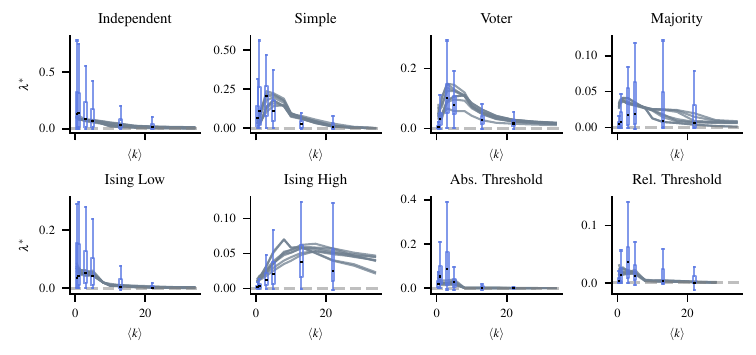}
    \caption{Asymptotic detectability $\lambda^*$ for each mechanism as a function of average degree $\langle k \rangle$. For eight spreading mechanisms $\mathbf{F}$ (each subplot), we systematically analyze a partition of the parameter space $\Omega|\mathbf{F}$ and obtain asymptotic detectability values $\lambda^*$. We obtain of points of a given asymptotic detectability. For each point, we simulate processes where we range system sizes over five orders of magnitude $N=10^1, \ldots, 10^5$, and obtain the True Positive Ratio on synthetic networks. For most mechanisms our results show  higher TPRs for higher asymptotic detectability and system size. These results, however, do not necessarily hold for higher $\lambda^*$ in the Voter model and the Majority Voter model. }
    \label{fig:ame-density}
\end{figure}

\begin{figure}
    \centering    
    \includegraphics[width=0.85\linewidth]{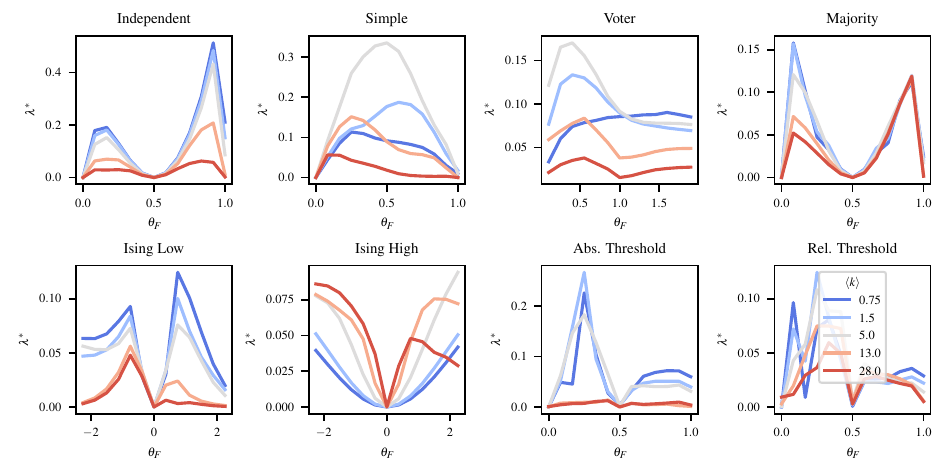}
    \caption{Relationship between topology and process parameters in detectability. Average asymptotic detectability $\lambda^*$ for each mechanism as a function of infection parameter $\theta_f$ and average degree $\langle k \rangle$ (colors). For most models, increasing network density results both in overall lower $\lambda^*$ and a shift in the more detectable $\theta_F$ values.}
    \label{fig:ame-thetak-detect}
\end{figure}

\begin{figure}
    \centering
    \includegraphics[width=\linewidth]{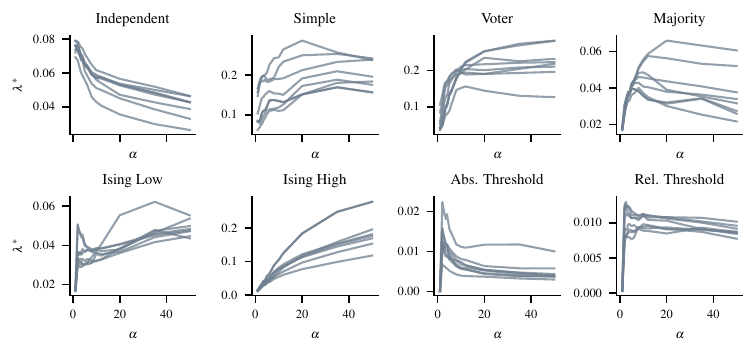}
    \caption{\textbf{Asymptotic detectability as a function of degree heterogeneity}. We assess the effect of degree heterogeneity on the detectability of eight spreading mechanisms. For eight spreading mechanisms $\mathbf{F}$ (each subplot), we systematically analyze a partition of the parameter space $\Omega|\mathbf{F}$ and obtain asymptotic detectability values $\lambda^*$. We obtain of points of a given asymptotic detectability. For each point, we simulate processes where we range system sizes over five orders of magnitude $N=10^1, \ldots, 10^5$, and obtain the True Positive Ratio on synthetic networks. For most mechanisms our results show  higher TPRs for higher asymptotic detectability and system size. These results, however, do not necessarily hold for higher $\lambda^*$ in the Voter model and the Majority Voter model. }
    \label{fig:si:param-space}
\end{figure}

\subsection{Alternative models: mechanisms of minimal divergence}
\label{sec:si:altmechs}

Our framework allows us to identify the best model alternatives; mechanisms that are prone to pairwise confounding. We do this by selecting the mechanism that minimizes the $\lambda^*$ per point in the parameter space. In Figure \ref{fig:altmechs} we depict the result of increasing the network density over the parameters $(\theta_F, \theta_R)$ of the SIS model, revealing a rich array of model alterative. The parameter region is characterized by a large presence of Voter, Independent and Majority Voter models. However, the regions where these models are present vary enough that escape a systematic characterization. Overall, the parameter spaces display a high fluidity of model alternatives, which can include up to six different models over different parameter values. 

Despite this large heterogeneity, aggregating results over $\langle k \rangle$ shows that the simple and voter models largely replace each other for sparse networks, but increasing the density --- at around $\langle k \rangle=10$--- increases the heterogeneity in best model alternatives. The independent and Ising models are largely dominated by the majority voter (we excluded different-temperature Ising models from this comparison, as they were trivially replaced by each other). The independent case can perhaps partly be explained as the majority voter includes a noise parameter that is functionally the independent model over a certain region of the dynamic range. 
Perhaps unsurprisingly, both threshold models tend to replace each other, although for sparser networks the relative version tends to be replaced by the voter model. 

We estimate the replacement ratio on synthetic networks, i.e., we characterize the ratio of models with higher likelihood than the true mechanism on simulations. To do this, we sample points from the model's parameter space $\Omega|\mathbf{F} = (\alpha, \mathbf{R}, \theta_F, \theta_R, i_0)$ and simulate processes on networks of varying system size $N$. These results differ from the asymptotic analysis as in these cases we perform statistical model selection on synthetic data and estimate the ratio on models that were misclassified. Replacement ratios depend on network size and  
For instance, when model selection fails for simple infections threshold models are likely to be selected. This is related to the mechanism's detectability within our parameter space: Threshold models are not the best alternative in general, but they are in higher-density networks, where model selection tends to fail. 

\begin{figure}[t]
    \centering
    \includegraphics[width=0.98\linewidth]{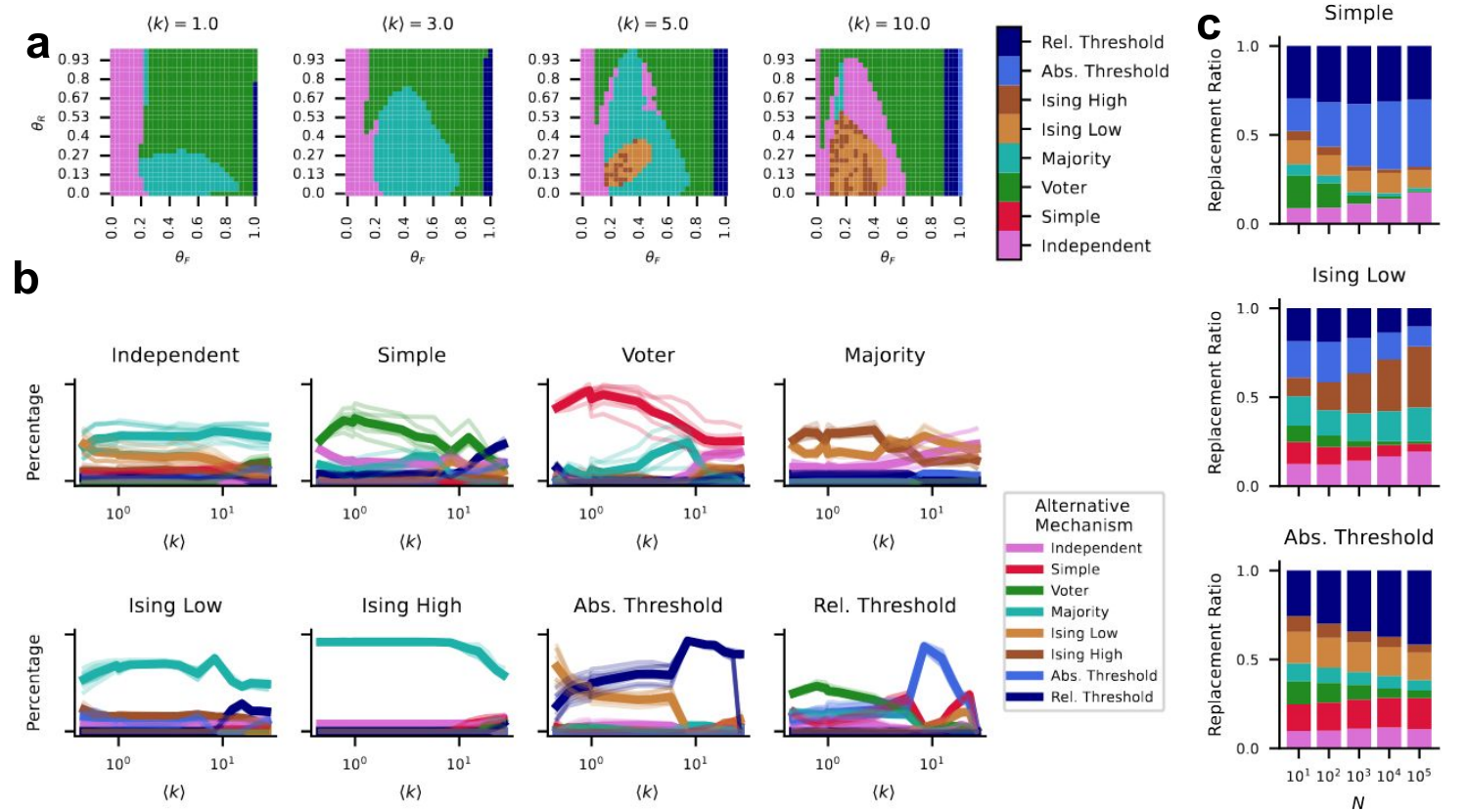}
    \caption{\textbf{Mechanisms alternatives with minimum divergence.} \textbf{a} Using AMEs we determine the mechanisms with minimum expected loglikelihood ratio. Best alternatives for Simple mechanism under SIS dynamics over its parameter space. Increasing density also increases the diversity of  alternative models. \textbf{b} Best alternatives averaged over the parameter space per mechanism. Percentage of the parameter space where each alternative mechanism had the smallest $\mathbb{E}[LR]$. \textbf{c} Replacement ratio of models given network size --- ratio of times an alternative mechanism was selected over the true model. }
    \label{fig:altmechs}
\end{figure}

\section{Empirical data}
\label{sec:si:emp_data}

We analyze five datasets of temporal dynamics on networks. Real-world data is rarely of the form of idealized models and thus require preprocessing regarding binarization of continuous feature dynamics and discretization of continuous or high-granularity time. 

\subsection{Main preprocessing algorithms}
\label{sec:si:preproc}

We preprocess each dataset according to their own features --- availability of different network topologies, features that serve as spreading processes, and characteristics of the data. Two common preprocessing steps consist of binarization of continuous features and discretization of continuous time. In pseudo-code, we use the following pipelines. 

\subsubsection*{Feature binarization}

We binarize features using a parameter $\phi$ that serves as a threshold for infection, e.g., a node is infected if it's current value is higher than $\phi$. We apply the same threshold to all nodes and account for node-specific variability by deseasoning trends and daily patterns, and standardizing deseasoned values. Algorithm \ref{alg:binarization} sketches the pseudocode use for feature binarization. We remove cyclical components to ensure that infected states reflect deviations not attributed to time (e.g., bike usage and traffic peak at certain hours), while we standardize values per node to ensure that deviations are comparable across nodes (e.g., some stations or avenues might have higher overall demand). 

\begin{algorithm}
\caption{Feature time series binarization}
\label{alg:binarization}
\begin{algorithmic}
\Require Time series of features per node $\mathbf{F}$, threshold $\phi$, cycle length $\eta$ 
\For{node $i = 1$ to $N$}
    \State Let $\{F_i(t)\}_t$ be the time series of feature values for node $i$
    \State Decompose $F_i$ into trend $T_i$, cyclical $C_i$ and error $E_i$ components:
    \[ F_i(t) = T_i(t) + C_i(t|\eta) + E_i(t) \]
    \State Remove seasonal and trend components to obtain deseasoned errors $E_i(t)$
    \State Obtain average errors and standard deviations over time \[\mu_i=\frac{1}{T}\sum_t E_i(t), \quad \sigma^2_i=\frac{1}{T}\sum_t (E_i(t)-\mu_i)^2\]
    \State Standardize errors, $
    \hat{E}_i(t) \gets \frac{E_i(t) - \mu_i}{\sigma_i}$
    \For{each time point $t$}
        \State $X_i(t) \gets \begin{cases} 1 & \text{if } \hat{E}_i(t) > \phi \\ 0 & \text{otherwise} \end{cases}$
    \EndFor
\EndFor
\State \Return Binary state matrix $\mathbf{X} \in \{0,1\}^{N \times T}$ where $X_i(t)$ is the binary indicator for node $i$ at time $t$
\end{algorithmic}
\end{algorithm}

\subsubsection*{Temporal discretization}

Unprocessed temporal data can be of high-granularity in a manner that resembles continuous time, e.g., milliseconds or minutes over long periods. We discretize temporal data over window sizes $\Delta t$ and assess it's impact on model selection. Temporal discretization requires determining rules for establishing the state of the window (e.g., for cases when there are multiple states changes within the same period). 

In the idealized scenarios we systematically analyze, the network is static and time does not have any units beyond the Markov property. We describe our datasets in terms of the different preprocessing requirements, as well as any potential violations of our assumptions. Only two datasets contain binary states from source (a game-theoretic study and engagement on an online discussion), while the others require dataset-specific definitions of binary states. Temporal aggregations also differ, with few allowing for a Markovian interpretation of time. This discretization directly affects our interpretation of the dynamics. While in many cases the states can be clearly defined (even if they require preprocessing), discretization defines the number of times a node is "susceptible". If a node has 1 infected neighbor for eleven seconds and then becomes infected, 1-second bins will result in 11 trials, but 10-second bins will result in two trials. In some datasets, we also consider cases when the data can be temporally split into different periods. For example, the dataset on bicycle usage includes a year and a half before and during the COVID pandemic. 

\begin{algorithm}
\caption{Discretization of continuous or high-granularity time}
\label{alg:discretization}
\begin{algorithmic}
\Require Node level time series data $\{\mathbf{D}_t\}_{t_0}^T$, observation window size $\Delta t$.
\State Set initial bounds $(\Gamma_0, \Gamma_1) \gets (t_0, t_0+\Delta t) $ and time $\tau \gets 0$
\While{$\Gamma_1 < T$}
    \For{node $i = 1$ to $N$}
    \State Set binary node state $X_i(\tau)$ for time $\tau$ (dataset-specific)
    \EndFor
    \State Increase bounds $(\Gamma_0, \Gamma_1)\gets(\Gamma_1, \Gamma_i+\Delta t)$ and time $\tau \gets \tau + 1$
    \EndWhile
\State \Return Binary state matrix $\mathbf{X}$
\end{algorithmic}
\end{algorithm}

\subsection{Description of datasets}
\label{sec:si:datasecs}

\textbf{Game}: A game-theoretic study performed on a computer game where participants had global and individual incentives to reach a consensus on a static network (a realization of an ER network). The global incentive was that any consensus would lead to a payoff, but individuals would get a larger payoff for particular states. Participants played simultaneously for a fixed amount of time to play, during which they had real-time access to the state of their neighbors (not their incentives), their neighbor's degree and connections among their neighbors. The available data includes timestamps of state changes, node degrees and a visualization of the topology, which was manually reconstructed (a potential source of errors that we except to be small). The end status reached a consensus.

\textit{Preprocessing}: this dataset closely aligns with our framework, including a static ER network and well-defined binary states. The main assumptions that are not met are temporal. While time is discrete with second-wide bins, we discretize time via a parameter $\Delta t$. The number of flips starts increasing once the time to reach a consensus approaches, with some users being stubborn on trying to "flip" the consensus to their preferred state.  

\textbf{Higgs}: Dataset on the spread of the news of the discovery of the Higgs Boson on an open social media platform. The data include a large sample of users with open follower information, i.e., both their followers and followees, which we use to reconstruct the underlying network. The dynamics correspond to activity regarding the discovery of the Higgs Boson. 

\textit{Preprocessing}: We define a user as "infected" if they participated in the conversation, defined through a collection of hashtags. A node is considered active for a duration $\Delta t$. Participation can be either retweeting, mentioning or commenting. 

\textbf{Urban}: Three datasets related to urban transportation --- two of average speed on road networks, and one of demand of public bicycles. For the bike network links correspond to physical proximity between stations, which we used from the original study. The dynamics for traffic networks correspond to average speed collected every five minutes, and for bike usage corresponds to the total demand of bikes aggregated every two hours. 

\textit{Preprocessing}: to define binary states, we first used a deseasonalizing technique at node-level to remove the effect variation within the day. This allowed us to identify deviations for the average daily behaviour, which we measure using standard deviations from the mean. For fine-grained speed data, we used the a $\Delta t$ aggregation period to obtain average speed during the observation period, and defined binary states using the same deseasonalizing technique. 

\textbf{Supply}: A supply-chain dataset within the factory of a large conglomerate in Bangladesh. Nodes correspond to products, and edges correspond to different relationships between products: whether they are produced in the same factory, stored in the same location, or belong to the same group. Each product contains four quantities: the amount produced, delivered, issued by the factory or the amount of orders by clients. For simplicity, we treat each quantity as different dynamics. Data is temporally aggregated per day. 

\textit{Preprocessing}: Using the same method as in Urban, we defined nodes as infected if their features (amounts produced, delivered, issued or ordered) deviated from average behaviour.   

\subsection{Additional results and data tables}
\label{sec:si:extrafit}

We include complementary results and fitted values ofr all our subdatasets. In Figure \ref{fig:emp_single} we depict the relative fraction of optimal \textit{single parameter} mechanisms when varying the primary parameter (either temporal granularity $\Delta t$ or binarization threshold $\phi$). In most cases, the independent model is the best fit, suggesting that an relevant part of the dynamics is not captured by topological data. However, as we show in the main text, including noisy variants that combine independent and another mechanism generally provides a better fit. The noisy versions are of the form $F^n_{k,m}=\theta_{nI}+(1-\theta_{nI})F_{k,m}(\theta_{nF})$ where $\theta_I$ is the independent parameter and $F_{k,m}(\theta_F)$ the mechanistic component.

\begin{figure}[t]
    \centering
    \includegraphics[width=0.98\linewidth]{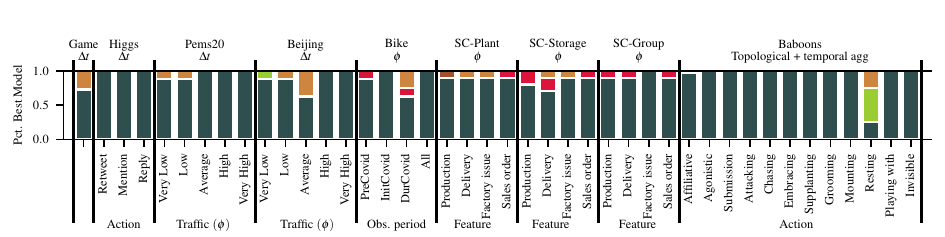}
    \caption{\textbf{Model selection for subdatasets without noisy model variants.} When fitting single infections, the most common model is Indepndent. We introduce the noisy variants to address this effect.}
    \label{fig:emp_single}
\end{figure}

As an example for parameter behavior in the Game and Higgs Retweet datasets, in Figure \ref{fig:theta} we include the maximum likelihood estimators for Independent and Simple parameters in their single and noisy components. For these datasets, the Independent component consistently displays exponential growth with $\Delta t$. For Higgs Retweet, the behavior of the Simple parameter differs when considered in single and noisy versions. 

\begin{figure}[t]
    \centering
    \includegraphics[width=0.5\linewidth]{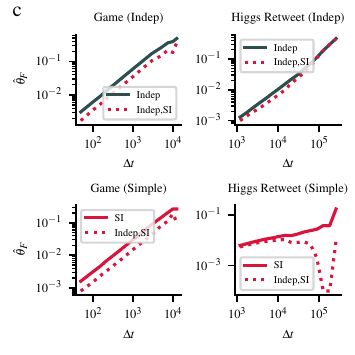}
    \caption{\textbf{MLE estimates for parameters as a function of $\Delta t$.} For (left) Game and (right) Higgs retweet we depict the Maximum Likelihood estimate $\hat{\theta}_F$ as a function of the temporal granularity $\Delta t$. Top rows depict the Independent parameter in single and noisy Simple versions, and bottom rows depict the Simple parameter in both single and noisy versions. The noisy parameter displays exponential growth with observation period, while the Simple parameter is dataset-dependent, while the noisy version collapses for the Higgs dataset.}
    \label{fig:theta}
\end{figure}

We provide complete results from our data fitting process. In Table \ref{tab:fullfit} we summarize the characteristics of each subdataset, including system size $N$, average degree $\langle k \rangle$, the value of the primary parameter and the total amount of data $S=\sum_t \sum_{k,m}S_{k,m}(t)$ given preprocessing. We include the optimal model abbreviated as independent I, simple S, voter V, majority M, Ising low IL, Ising high IH, absolute threshold TA and relative threshold TR. In all cases, if the noisy version is the optimal parameter we include the prefix \textit{n} (e.g., nS for noisy simple infections). Since, in many cases, the independent model is the optimal fit with single mechanisms, we include the loglikelihood of the independent model as well as the estimated infection parameter $\hat{\theta}_I$. We also include the loglikelihood of the optimal model; for the cases where the optimal is a noisy variant, we include the estimated parameters for both it's components.

{\tiny
\begin{center}
\begin{longtable}{|c|c|c|c|c|c|c|c|c|c|c|c|}
\caption{Results from empirical data analysis. For each dataset and subdataset we include the system size $N$, average degree $\langle k \rangle$, total amount of susceptible nodes $S$, value of primary parameter \textit{PPv}, optimal \textit{Model} (prefix \textit{n} for noisy versions), loglikelihood of indpendent model \textit{llog I}, loglikelihood of optimal model \textit{llog opt}, estimated independent infection rate $\hat{\theta}_I$, estimated independent component for optimal noisy model $\hat{\theta}_{nI}$ and estimated non-independent component for optimal noisy model $\hat{\theta}_{nF}$} \label{tab:fullfit} \\

\hline \multicolumn{1}{|c|}{\textbf{Dataset}} & \multicolumn{1}{c|}{\textbf{Subdataset}} & \multicolumn{1}{c|}{$N$} &\multicolumn{1}{c|}{$\langle k \rangle$} & \multicolumn{1}{c|}{$S$} & \multicolumn{1}{c|}{\textbf{PPv}} & \multicolumn{1}{c|}{\textbf{Model}}& \multicolumn{1}{c|}{\textbf{llog I}}& \multicolumn{1}{c|}{\textbf{llog opt}}& \multicolumn{1}{c|}{$\hat{\theta}_I$}& \multicolumn{1}{c|}{$\hat{\theta}_{nI}$}& \multicolumn{1}{c|}{$\hat{\theta}_{nF}$} \\ \hline 
\endfirsthead

\multicolumn{12}{c}%
{{\bfseries \tablename\ \thetable{} -- continued from previous page}} \\
\hline \multicolumn{1}{|c|}{\textbf{Dataset}} & \multicolumn{1}{c|}{\textbf{Subdataset}} & \multicolumn{1}{c|}{$N$} &\multicolumn{1}{c|}{$\langle k \rangle$} & \multicolumn{1}{c|}{$S$} & \multicolumn{1}{c|}{\textbf{PPv}} & \multicolumn{1}{c|}{\textbf{Model}}& \multicolumn{1}{c|}{\textbf{llog I}}& \multicolumn{1}{c|}{\textbf{llog opt}}& \multicolumn{1}{c|}{$\hat{\theta}_I$}& \multicolumn{1}{c|}{$\hat{\theta}_{nI}$}& \multicolumn{1}{c|}{$\hat{\theta}_{nF}$} \\ \hline 
\endhead

\hline \multicolumn{12}{|r|}{{Continued on next page}} \\ \hline
\endfoot

\hline \hline
\endlastfoot

game&&36&5.556&19408.0&50&nS&-586.6762&-580.8917&0.0031&0.0017&0.0008\\
game&&36&5.556&9721.0&100&nS&-519.3507&-514.3975&0.0061&0.0035&0.0015\\
game&&36&5.556&6478.0&150&nS&-477.895&-472.6751&0.0089&0.0051&0.0023\\
game&&36&5.556&4856.0&200&nS&-460.0115&-455.1623&0.0121&0.0071&0.003\\
game&&36&5.556&1942.0&500&nS&-376.2042&-371.4125&0.0299&0.0175&0.0073\\
game&&36&5.556&977.0&1000&nS&-317.4475&-312.6133&0.0594&0.0349&0.0148\\
game&&36&5.556&336.0&3000&nTA&-222.9901&-216.9985&0.1726&0.1672&0.838\\
game&&36&5.556&228.0&5000&nS&-186.5396&-180.2081&0.2544&0.1455&0.0793\\
game&&36&5.556&161.0&7500&nTA&-151.8052&-27.368&0.3602&0.0001&0.0072\\
game&&36&5.556&148.0&10000&IL&-142.9694&-126.3647&0.3919&&0.1332\\
game&&36&5.556&120.0&12500&nIL&-119.9038&-115.6422&0.4833&0.0585&0.0779\\
higgs&Retweet&456626&16.267&228913053.0&1200&nS&-3290711.0811&-3152565.3879&0.0013&0.001&0.0054\\
higgs&Retweet&456626&16.267&159966156.0&1717&nS&-3099026.5149&-2954514.4103&0.0018&0.0014&0.0063\\
higgs&Retweet&456626&16.267&111111220.0&2457&nS&-2909583.371&-2759987.6451&0.0026&0.0019&0.0072\\
higgs&Retweet&456626&16.267&77324996.0&3516&nS&-2725161.7212&-2574284.2538&0.0037&0.0025&0.0081\\
higgs&Retweet&456626&16.267&54040877.0&5031&nS&-2545831.2509&-2402903.3092&0.0052&0.0035&0.0089\\
higgs&Retweet&456626&16.267&37149940.0&7200&nS&-2361024.9405&-2230203.5592&0.0075&0.0049&0.0095\\
higgs&Retweet&456626&16.267&25738951.0&10302&nS&-2184775.4492&-2071257.0055&0.0106&0.0069&0.0101\\
higgs&Retweet&456626&16.267&17524567.0&14743&nS&-2001421.1389&-1899227.3311&0.0153&0.0099&0.0107\\
higgs&Retweet&456626&16.267&12052023.0&21097&nS&-1829581.8773&-1778282.48&0.0219&0.0165&0.008\\
higgs&Retweet&456626&16.267&8402888.0&30189&nS&-1666032.8828&-1624539.5013&0.0308&0.0235&0.0082\\
higgs&Retweet&456626&16.267&5215081.0&43200&nS&-1441499.8789&-1409050.9809&0.0477&0.0366&0.0085\\
higgs&Retweet&456626&16.267&3396257.0&61817&nS&-1249332.41&-1231619.3058&0.0705&0.0582&0.0069\\
higgs&Retweet&456626&16.267&2047365.0&88459&nS&-1036139.2041&-1034663.1527&0.112&0.1076&0.002\\
higgs&Retweet&456626&16.267&1171761.0&126582&I&-818373.0668&-818373.0668&0.1885&&0.1885\\
higgs&Retweet&456626&16.267&730109.0&181135&I&-638221.4486&-638221.4486&0.2942&&0.2942\\
higgs&Retweet&456626&16.267&432350.0&259199&nS&-430387.1036&-429874.9753&0.4604&0.4504&0.0093\\
higgs&Mention&456626&16.267&229088636.0&1200&nS&-1599699.3913&-1510767.4397&0.0006&0.0004&0.0031\\
higgs&Mention&456626&16.267&160139748.0&1717&nS&-1514564.0964&-1422185.8892&0.0008&0.0005&0.0035\\
higgs&Mention&456626&16.267&111739135.0&2457&nS&-1430937.8879&-1337476.8659&0.0011&0.0007&0.0039\\
higgs&Mention&456626&16.267&77494115.0&3516&nS&-1347682.8462&-1254855.9453&0.0016&0.001&0.0044\\
higgs&Mention&456626&16.267&54207868.0&5031&nS&-1268260.8662&-1179371.2301&0.0023&0.0014&0.0048\\
higgs&Mention&456626&16.267&37314592.0&7200&nS&-1186232.6078&-1103297.2033&0.0033&0.0019&0.0052\\
higgs&Mention&456626&16.267&25900760.0&10302&nS&-1108483.6301&-1033023.0242&0.0046&0.0026&0.0056\\
higgs&Mention&456626&16.267&18139393.0&14743&nS&-1034244.4157&-967244.7462&0.0066&0.0037&0.0058\\
higgs&Mention&456626&16.267&12207296.0&21097&nS&-952265.3802&-909021.5689&0.0096&0.0061&0.0053\\
higgs&Mention&456626&16.267&8555730.0&30189&nS&-881116.9639&-848490.1958&0.0135&0.009&0.005\\
higgs&Mention&456626&16.267&5364195.0&43200&nS&-775615.1987&-745538.7292&0.0206&0.0132&0.0056\\
higgs&Mention&456626&16.267&3541659.0&61817&nS&-689714.103&-665669.2047&0.0301&0.0195&0.0059\\
higgs&Mention&456626&16.267&2182221.0&88459&nS&-592934.949&-584028.8231&0.0466&0.0368&0.0043\\
higgs&Mention&456626&16.267&1284640.0&126582&nS&-496784.1887&-493085.6724&0.0757&0.0657&0.0035\\
higgs&Mention&456626&16.267&834338.0&181135&nS&-425181.496&-425138.9496&0.1132&0.1121&0.0003\\
higgs&Mention&456626&16.267&443333.0&259199&nS&-314449.064&-307043.95&0.1938&0.1521&0.0229\\
higgs&Reply&456626&16.267&229192182.0&1200&nS&-462588.8534&-451393.6592&0.0001&0.0001&0.0015\\
higgs&Reply&456626&16.267&160242039.0&1717&nS&-441010.913&-428957.9527&0.0002&0.0002&0.0017\\
higgs&Reply&456626&16.267&111840098.0&2457&nS&-420396.6363&-407268.0029&0.0003&0.0002&0.002\\
higgs&Reply&456626&16.267&77593596.0&3516&nS&-399323.4881&-385514.5931&0.0004&0.0003&0.0024\\
higgs&Reply&456626&16.267&54306003.0&5031&nS&-379510.6029&-365224.3888&0.0006&0.0004&0.0027\\
higgs&Reply&456626&16.267&37411279.0&7200&nS&-359276.0663&-345234.6205&0.0008&0.0006&0.003\\
higgs&Reply&456626&16.267&25996044.0&10302&nS&-339899.2139&-326474.0239&0.0012&0.0009&0.0033\\
higgs&Reply&456626&16.267&18233695.0&14743&nS&-321416.78&-309551.1411&0.0016&0.0012&0.0034\\
higgs&Reply&456626&16.267&12298322.0&21097&nS&-300827.7975&-292261.2273&0.0024&0.0019&0.0032\\
higgs&Reply&456626&16.267&8645577.0&30189&nS&-283062.4697&-275487.9335&0.0034&0.0026&0.0033\\
higgs&Reply&456626&16.267&5450200.0&43200&nS&-256569.6002&-249665.2538&0.0052&0.0039&0.0037\\
higgs&Reply&456626&16.267&3624542.0&61817&nS&-235031.7571&-227672.5965&0.0076&0.0053&0.0046\\
higgs&Reply&456626&16.267&2257033.0&88459&nS&-210224.178&-207248.3987&0.0119&0.0095&0.0034\\
higgs&Reply&456626&16.267&1347750.0&126582&nS&-185057.8466&-183802.1239&0.0193&0.0167&0.0026\\
higgs&Reply&456626&16.267&892696.0&181135&nS&-166492.8069&-166435.2993&0.0284&0.0278&0.0005\\
higgs&Reply&456626&16.267&453650.0&259199&nS&-133198.3288&-130465.6989&0.0517&0.0407&0.0176\\
Pems20&Very low (-2)&325&12.794&967283.0&300&nTA&-345659.987&-210561.3478&0.0677&0.2561&0.0166\\
Pems20&Very low (-2)&325&12.794&144563.0&1800&I&-102576.4645&-102576.4645&0.1939&&0.1939\\
Pems20&Very low (-2)&325&12.794&72217.0&3600&nTA&-61083.2826&-47324.5636&0.2731&0.2978&0.2509\\
Pems20&Very low (-2)&325&12.794&37899.0&7200&nTA&-37052.6656&-33049.8124&0.4123&0.4261&0.6484\\
Pems20&Very low (-2)&325&12.794&14734.0&21600&nTA&-12839.7342&-101.1254&0.7079&0.9921&0.0125\\
Pems20&Very low (-2)&325&12.794&3764.0&86400&nTA&-3275.5839&-196.3387&0.7088&0.9771&0.115\\
Pems20&Very low (-2)&325&12.794&2431.0&129600&nTA&-2113.1646&-82.1096&0.7096&0.9663&0.205\\
Pems20&Very low (-2)&325&12.794&2186.0&172800&nTA&-1741.8726&-104.9436&0.7589&0.9714&0.3361\\
Pems20&Low (1)&325&12.794&1560043.0&300&nS&-772171.8368&-758197.4252&0.1084&0.0551&0.0077\\
Pems20&Low (1)&325&12.794&212177.0&1800&nS&-171295.4395&-171284.8124&0.2475&0.2423&0.0008\\
Pems20&Low (1)&325&12.794&109377.0&3600&nTA&-100200.7605&-95010.0687&0.3312&0.461&0.4082\\
Pems20&Low (1)&325&12.794&59762.0&7200&nTA&-59373.3676&-41161.3958&0.4526&0.5323&0.2831\\
Pems20&Low (1)&325&12.794&26168.0&21600&nTA&-19567.6376&-5921.9037&0.7867&0.4928&0.1258\\
Pems20&Low (1)&325&12.794&6755.0&86400&nTA&-5380.1226&-2509.5196&0.7591&0.5832&0.4074\\
Pems20&Low (1)&325&12.794&5841.0&129600&nTA&-3344.4425&-615.786&0.8644&0.9759&0.3916\\
Pems20&Low (1)&325&12.794&4153.0&172800&nTA&-2706.1119&-731.8148&0.8327&0.7426&0.2512\\
Pems20&Average (0)&325&12.794&5975360.0&300&nS&-3461572.8489&-3289418.3175&0.1381&0.0492&0.0181\\
Pems20&Average (0)&325&12.794&1779161.0&1800&nS&-984537.5723&-951406.3117&0.1285&0.0727&0.0189\\
Pems20&Average (0)&325&12.794&895420.0&3600&nTA&-577006.0216&-232751.7885&0.1643&0.3925&0.0102\\
Pems20&Average (0)&325&12.794&449589.0&7200&nS&-353635.6015&-349925.8343&0.235&0.1921&0.0176\\
Pems20&Average (0)&325&12.794&148844.0&21600&nS&-148783.0559&-148775.6566&0.4881&0.4845&0.0021\\
Pems20&Average (0)&325&12.794&34458.0&86400&nIL&-34168.9896&-34167.4978&0.5539&0.1041&-0.0013\\
Pems20&Average (0)&325&12.794&22775.0&129600&nIL&-15020.8999&-14529.9463&0.8292&0.5612&-0.0472\\
Pems20&Average (0)&325&12.794&16850.0&172800&nTA&-16177.8677&-15657.3545&0.617&0.418&0.1743\\
Pems20&High (1)&325&12.794&19650431.0&300&nS&-951755.954&-926335.5501&0.0054&0.0041&0.0071\\
Pems20&High (1)&325&12.794&2978808.0&1800&nS&-602592.4734&-539495.2826&0.0316&0.0137&0.0236\\
Pems20&High (1)&325&12.794&1480682.0&3600&nS&-403022.0716&-377953.3484&0.0467&0.0278&0.0249\\
Pems20&High (1)&325&12.794&732387.0&7200&nS&-281689.3522&-276533.616&0.0751&0.0605&0.0188\\
Pems20&High (1)&325&12.794&237694.0&21600&nS&-131212.4766&-130617.5165&0.128&0.1177&0.0121\\
Pems20&High (1)&325&12.794&62597.0&86400&nS&-25246.4636&-24874.3381&0.0803&0.0676&0.0199\\
Pems20&High (1)&325&12.794&40810.0&129600&nS&-21371.6027&-21037.8184&0.118&0.1011&0.0239\\
Pems20&High (1)&325&12.794&30585.0&172800&nS&-14827.8196&-14347.5823&0.105&0.0811&0.0358\\
Pems20&Very High (2)&325&12.794&19932208.0&300&nS&-27004.3757&-26790.8654&0.0001&0.0001&0.0013\\
Pems20&Very High (2)&325&12.794&3295279.0&1800&nS&-61421.8809&-58407.521&0.0017&0.0014&0.0096\\
Pems20&Very High (2)&325&12.794&1647095.0&3600&nS&-39435.9255&-37894.4063&0.0023&0.0019&0.0101\\
Pems20&Very High (2)&325&12.794&822614.0&7200&nS&-27340.7241&-26857.5668&0.0034&0.003&0.0081\\
Pems20&Very High (2)&325&12.794&272089.0&21600&nS&-23806.5664&-23790.7974&0.011&0.0108&0.002\\
Pems20&Very High (2)&325&12.794&66089.0&86400&nS&-14921.2972&-14712.2977&0.0364&0.0318&0.0159\\
Pems20&Very High (2)&325&12.794&44485.0&129600&nS&-8684.8523&-8589.2716&0.0302&0.027&0.0137\\
Pems20&Very High (2)&325&12.794&32702.0&172800&nS&-8259.2759&-8231.6195&0.0423&0.0398&0.0078\\
Beijing&Very Low (-2)&3126&2.146&3190023.0&300&nV&-1674190.0278&-1290219.1074&0.1184&0.0397&1.95\\
Beijing&Very Low (-2)&3126&2.146&342935.0&1800&nV&-341592.171&-297634.4798&0.5368&0.448&1.9318\\
Beijing&Very Low (-2)&3126&2.146&161275.0&3600&nV&-137917.8405&-125383.9947&0.7202&0.6759&1.95\\
Beijing&Very Low (-2)&3126&2.146&87061.0&7200&nTA&-48250.8193&-27134.3746&0.8712&0.7509&0.0001\\
Beijing&Very Low (-2)&3126&2.146&28674.0&21600&nTA&-1097.0901&-429.4406&0.9959&0.9947&0.0001\\
Beijing&Very Low (-2)&3126&2.146&9041.0&86400&nTA&-160.1531&-77.1069&0.9984&0.9974&0.0001\\
Beijing&Very Low (-2)&3126&2.146&3107.0&129600&nTA&-44.1723&-22.1087&0.9987&0.9987&0.0313\\
Beijing&Very Low (-2)&3126&2.146&4089.0&172800&nTA&-134.8256&-107.8753&0.9966&0.9894&0.0001\\
Beijing&Low (-1)&3126&2.146&6382280.0&300&nV&-3286279.9342&-2460540.2538&0.115&0.0396&1.95\\
Beijing&Low (-1)&3126&2.146&886743.0&1800&nV&-882650.6949&-761128.5678&0.46&0.3615&1.8512\\
Beijing&Low (-1)&3126&2.146&447064.0&3600&nV&-440993.2827&-396768.3676&0.5685&0.5031&1.95\\
Beijing&Low (-1)&3126&2.146&265410.0&7200&nTA&-237858.8673&-148044.6533&0.6874&0.6608&0.0496\\
Beijing&Low (-1)&3126&2.146&117123.0&21600&nTA&-22598.5629&-12632.0111&0.9703&0.954&0.0001\\
Beijing&Low (-1)&3126&2.146&26441.0&86400&nV&-6121.0795&-5597.7565&0.9623&0.9595&1.95\\
Beijing&Low (-1)&3126&2.146&21363.0&129600&nV&-2606.3363&-2463.5194&0.9834&0.9814&1.95\\
Beijing&Low (-1)&3126&2.146&15468.0&172800&nV&-3121.1649&-2927.7958&0.9685&0.9663&1.95\\
Beijing&Average (0)&3126&2.146&21662610.0&300&nV&-10516242.9687&-8162980.4674&0.1053&0.0433&1.95\\
Beijing&Average (0)&3126&2.146&4870823.0&1800&nTA&-4067252.1565&-2487066.7775&0.2656&0.173&0.0323\\
Beijing&Average (0)&3126&2.146&2474637.0&3600&nTA&-2198666.9962&-1565263.8883&0.306&0.3803&0.0663\\
Beijing&Average (0)&3126&2.146&1260282.0&7200&nV&-1196920.1804&-1134666.8698&0.3688&0.3294&1.95\\
Beijing&Average (0)&3126&2.146&421439.0&21600&nV&-403966.9232&-395532.5116&0.6193&0.6059&1.95\\
Beijing&Average (0)&3126&2.146&108512.0&86400&nTA&-100737.4923&-68926.0697&0.6563&0.6718&0.0531\\
Beijing&Average (0)&3126&2.146&76321.0&129600&nTA&-58731.9167&-41212.8843&0.7748&0.6224&0.0001\\
Beijing&Average (0)&3126&2.146&53793.0&172800&nV&-49564.1873&-47960.665&0.6635&0.6463&1.95\\
Beijing&High (1)&3126&2.146&66001533.0&300&nS&-4157460.7235&-3783839.9762&0.0074&0.0052&0.0707\\
Beijing&High (1)&3126&2.146&10244757.0&1800&nV&-2240135.5892&-2097816.0878&0.0349&0.0273&1.95\\
Beijing&High (1)&3126&2.146&5112493.0&3600&nV&-1353282.3939&-1304014.3682&0.045&0.0381&1.95\\
Beijing&High (1)&3126&2.146&2500367.0&7200&nV&-911765.6011&-897250.4103&0.0697&0.0625&1.95\\
Beijing&High (1)&3126&2.146&810104.0&21600&I&-445642.9172&-445642.9172&0.1273&&0.1273\\
Beijing&High (1)&3126&2.146&200663.0&86400&I&-112491.2546&-112491.2546&0.1311&&0.1311\\
Beijing&High (1)&3126&2.146&133421.0&129600&I&-80562.4921&-80562.4921&0.1476&&0.1476\\
Beijing&High (1)&3126&2.146&97195.0&172800&I&-60129.2915&-60129.2915&0.1535&&0.1535\\
Beijing&Very High (2)&3126&2.146&67496101.0&300&nS&-158114.188&-156326.2199&0.0002&0.0002&0.0091\\
Beijing&Very High (2)&3126&2.146&11004300.0&1800&nS&-488958.9999&-458567.9742&0.0049&0.0041&0.1122\\
Beijing&Very High (2)&3126&2.146&5496751.0&3600&nS&-342190.3239&-334298.7786&0.0073&0.0067&0.08\\
Beijing&Very High (2)&3126&2.146&2742167.0&7200&nS&-216716.7611&-214990.1242&0.0097&0.0094&0.0463\\
Beijing&Very High (2)&3126&2.146&911072.0&21600&nS&-84993.5283&-84963.3644&0.0119&0.0119&0.0088\\
Beijing&Very High (2)&3126&2.146&221587.0&86400&I&-42830.6669&-42830.6669&0.0298&&0.0298\\
Beijing&Very High (2)&3126&2.146&149565.0&129600&I&-27020.5607&-27020.5607&0.0273&&0.0273\\
Beijing&Very High (2)&3126&2.146&108982.0&172800&I&-22823.6482&-22823.6482&0.033&&0.033\\
Bike&Pre-covid&112&6.857&29583.0&-1.0&nTA&-4616.7667&-0.0&0.0226&0.938&0.0229\\
Bike&Pre-covid&112&6.857&106982.0&-0.5&nTA&-79311.5036&-71371.2099&0.2099&0.2051&0.2531\\
Bike&Pre-covid&112&6.857&180550.0&0.0&nTA&-107030.8973&-84745.8843&0.1433&0.105&0.4082\\
Bike&Pre-covid&112&6.857&202061.0&0.5&nS&-101601.1848&-95166.4406&0.111&0.0683&0.0659\\
Bike&Pre-covid&112&6.857&216117.0&1.0&nS&-89547.8341&-84066.0018&0.0835&0.053&0.0618\\
Bike&Pre-covid&112&6.857&226853.0&1.5&nTA&-72402.8228&-68358.141&0.0579&0.075&0.4082\\
Bike&Pre-covid&112&6.857&234173.0&2.0&nS&-54868.4074&-52241.384&0.0383&0.027&0.0462\\
Bike&Pre-covid&112&6.857&238785.0&2.5&nS&-40095.8385&-38609.8121&0.0248&0.0189&0.0368\\
Bike&Init. Covid&112&6.857&80549.0&-0.5&nS&-56654.5794&-54641.8875&0.1909&0.0904&0.0387\\
Bike&Init. Covid&112&6.857&190993.0&0.0&nS&-98440.3738&-90812.6864&0.1152&0.0638&0.0711\\
Bike&Init. Covid&112&6.857&209476.0&0.5&nS&-83748.8503&-75971.3111&0.0793&0.0429&0.0711\\
Bike&Init. Covid&112&6.857&221286.0&1.0&nS&-67283.6281&-60088.1874&0.0542&0.029&0.0719\\
Bike&Init. Covid&112&6.857&227725.0&1.5&nS&-55157.6476&-49044.6793&0.04&0.0219&0.0704\\
Bike&Init. Covid&112&6.857&232425.0&2.0&nS&-43592.3485&-38763.8016&0.0286&0.0167&0.0658\\
Bike&Init. Covid&112&6.857&235667.0&2.5&nS&-34802.5795&-30966.201&0.0211&0.0128&0.0617\\
Bike&Dur. covid&112&6.857&8330.0&-1.0&nTA&-4436.5238&-0.0&0.1211&0.9527&0.3974\\
Bike&Dur. covid&112&6.857&93756.0&-0.5&nS&-76990.7982&-73500.0743&0.2563&0.1494&0.0593\\
Bike&Dur. covid&112&6.857&170974.0&0.0&nS&-107346.1117&-96458.4882&0.1573&0.077&0.0882\\
Bike&Dur. covid&112&6.857&197001.0&0.5&nS&-101595.3736&-89024.1477&0.1153&0.0511&0.0974\\
Bike&Dur. covid&112&6.857&213669.0&1.0&nS&-87786.1408&-78199.4404&0.0825&0.042&0.0892\\
Bike&Dur. covid&112&6.857&225252.0&1.5&nS&-70901.4922&-63953.8733&0.0568&0.0325&0.0766\\
Bike&Dur. covid&112&6.857&232656.0&2.0&nS&-56066.6685&-50544.2277&0.0397&0.0235&0.0693\\
Bike&Dur. covid&112&6.857&238241.0&2.5&nTA&-40440.5044&-26078.4636&0.0252&0.0161&0.0302\\
Bike&Full&112&6.857&230136.0&-0.5&nS&-187145.2071&-179677.1977&0.2512&0.1348&0.054\\
Bike&Full&112&6.857&547965.0&0.0&nTA&-304006.8365&-244979.6025&0.129&0.0594&0.408\\
Bike&Full&112&6.857&613857.0&0.5&nS&-276473.256&-247343.9923&0.0942&0.0479&0.0799\\
Bike&Full&112&6.857&655992.0&1.0&nS&-236875.8287&-211488.6934&0.0687&0.0359&0.0787\\
Bike&Full&112&6.857&682158.0&1.5&nS&-197296.6157&-177759.674&0.0507&0.0286&0.0722\\
Bike&Full&112&6.857&701872.0&2.0&nS&-154873.2828&-140746.929&0.0354&0.0214&0.0635\\
Bike&Full&112&6.857&715523.0&2.5&nS&-115926.9787&-106133.0785&0.0237&0.0153&0.0551\\
SC-Plant&Production&40&18.0&2357.0&-2.0&nTA&-280.6608&-122.0987&0.0161&0.1562&0.697\\
SC-Plant&Production&40&18.0&3477.0&-1.5&nTA&-1718.6032&-0.0&0.1081&0.4195&0.1298\\
SC-Plant&Production&40&18.0&5588.0&-1.0&nS&-3471.6838&-3471.6398&0.1546&0.1522&0.0004\\
SC-Plant&Production&40&18.0&6714.0&-0.5&nS&-3759.5524&-3746.7964&0.1309&0.1064&0.0065\\
SC-Plant&Production&40&18.0&7417.0&0.0&nS&-3367.2243&-3334.3216&0.0953&0.0676&0.0107\\
SC-Plant&Production&40&18.0&7941.0&0.5&nS&-2812.1145&-2784.7249&0.0669&0.0502&0.0098\\
SC-Plant&Production&40&18.0&8315.0&1.0&nS&-2093.7651&-2069.6141&0.0421&0.032&0.0102\\
SC-Plant&Production&40&18.0&8512.0&1.5&nS&-1499.6157&-1472.294&0.0264&0.019&0.0124\\
SC-Plant&Production&40&18.0&8658.0&2.0&nS&-894.1557&-883.7942&0.0135&0.011&0.0083\\
SC-Plant&Production&40&18.0&8729.0&2.5&nS&-559.9726&-559.966&0.0076&0.0075&0.0002\\
SC-Plant&Delivery&40&18.0&2215.0&-2.0&nTA&-42.2179&-0.0&0.0018&0.4308&0.3079\\
SC-Plant&Delivery&40&18.0&2360.0&-1.5&nTA&-418.985&-135.7208&0.0267&0.4921&0.4379\\
SC-Plant&Delivery&40&18.0&3145.0&-1.0&nTA&-1975.1713&-627.9379&0.1574&0.3665&0.1406\\
SC-Plant&Delivery&40&18.0&5295.0&-0.5&I&-3842.5243&-3842.5243&0.2019&&0.2019\\
SC-Plant&Delivery&40&18.0&6462.0&0.0&nS&-4055.77&-4051.7748&0.1572&0.1427&0.0037\\
SC-Plant&Delivery&40&18.0&7353.0&0.5&nS&-3697.4575&-3687.8469&0.111&0.0965&0.0057\\
SC-Plant&Delivery&40&18.0&7969.0&1.0&nS&-2875.4661&-2859.2775&0.0686&0.0576&0.0072\\
SC-Plant&Delivery&40&18.0&8344.0&1.5&nS&-2059.3144&-2042.1227&0.041&0.0338&0.0081\\
SC-Plant&Delivery&40&18.0&8555.0&2.0&nS&-1347.2811&-1331.8599&0.0229&0.0185&0.0089\\
SC-Plant&Delivery&40&18.0&8676.0&2.5&nS&-812.9044&-798.2173&0.012&0.0095&0.0098\\
SC-Plant&Factory Issue&40&18.0&2371.0&-1.5&nTA&-485.0832&-0.0&0.032&0.6996&0.1101\\
SC-Plant&Factory Issue&40&18.0&3741.0&-1.0&I&-2383.5457&-2383.5457&0.1612&&0.1612\\
SC-Plant&Factory Issue&40&18.0&5560.0&-0.5&I&-3829.363&-3829.363&0.184&&0.184\\
SC-Plant&Factory Issue&40&18.0&6828.0&0.0&nS&-3984.1511&-3972.3337&0.1397&0.1179&0.0065\\
SC-Plant&Factory Issue&40&18.0&7596.0&0.5&nS&-3425.5041&-3407.8903&0.0944&0.0782&0.0073\\
SC-Plant&Factory Issue&40&18.0&8078.0&1.0&nS&-2732.9244&-2714.8086&0.0628&0.0515&0.008\\
SC-Plant&Factory Issue&40&18.0&8417.0&1.5&nS&-1887.0755&-1870.835&0.0361&0.0291&0.0089\\
SC-Plant&Factory Issue&40&18.0&8579.0&2.0&nS&-1320.8948&-1307.9463&0.0223&0.0186&0.0082\\
SC-Plant&Factory Issue&40&18.0&8679.0&2.5&nS&-838.3075&-834.9408&0.0124&0.0113&0.0045\\
SC-Plant&Sales Order&40&18.0&1563.0&-1.5&nTA&-793.5674&-0.0&0.1126&0.8595&0.0599\\
SC-Plant&Sales Order&40&18.0&3712.0&-1.0&I&-2818.7441&-2818.7441&0.2196&&0.2196\\
SC-Plant&Sales Order&40&18.0&5796.0&-0.5&I&-4461.6292&-4461.6292&0.2253&&0.2253\\
SC-Plant&Sales Order&40&18.0&6912.0&0.0&I&-4410.7594&-4410.7594&0.1616&&0.1616\\
SC-Plant&Sales Order&40&18.0&7642.0&0.5&I&-3727.0848&-3727.0848&0.106&&0.106\\
SC-Plant&Sales Order&40&18.0&8118.0&1.0&I&-2833.4219&-2833.4219&0.0655&&0.0655\\
SC-Plant&Sales Order&40&18.0&8402.0&1.5&I&-2081.0038&-2081.0038&0.0412&&0.0412\\
SC-Plant&Sales Order&40&18.0&8578.0&2.0&nS&-1396.5138&-1395.4911&0.0239&0.0232&0.0018\\
SC-Plant&Sales Order&40&18.0&8677.0&2.5&nS&-900.715&-897.208&0.0136&0.0126&0.0041\\
SC-Storage&Production&40&33.25&2357.0&-2.0&nTA&-280.6608&-0.0&0.0161&0.8873&0.5738\\
SC-Storage&Production&40&33.25&3477.0&-1.5&nTA&-1718.6032&-0.0&0.1081&0.2427&0.0748\\
SC-Storage&Production&40&33.25&5588.0&-1.0&nS&-3471.6838&-3442.4728&0.1546&0.0776&0.0071\\
SC-Storage&Production&40&33.25&6714.0&-0.5&nS&-3759.5524&-3715.7077&0.1309&0.0696&0.0083\\
SC-Storage&Production&40&33.25&7417.0&0.0&nS&-3367.2243&-3320.618&0.0953&0.0505&0.0089\\
SC-Storage&Production&40&33.25&7941.0&0.5&nS&-2812.1145&-2780.4151&0.0669&0.0436&0.0071\\
SC-Storage&Production&40&33.25&8315.0&1.0&nS&-2093.7651&-2063.3028&0.0421&0.0261&0.0083\\
SC-Storage&Production&40&33.25&8512.0&1.5&nS&-1499.6157&-1474.1463&0.0264&0.017&0.0081\\
SC-Storage&Production&40&33.25&8658.0&2.0&nS&-894.1557&-882.3036&0.0135&0.0098&0.0064\\
SC-Storage&Production&40&33.25&8729.0&2.5&nS&-559.9726&-559.8994&0.0076&0.0074&0.0005\\
SC-Storage&Delivery&40&33.25&2215.0&-2.0&nIL&-42.2179&-0.0&0.0018&0.7286&2.2272\\
SC-Storage&Delivery&40&33.25&2360.0&-1.5&nTA&-418.985&-0.0&0.0267&0.2416&0.0751\\
SC-Storage&Delivery&40&33.25&3145.0&-1.0&nTA&-1975.1713&-578.276&0.1574&0.6019&0.2595\\
SC-Storage&Delivery&40&33.25&5295.0&-0.5&nTA&-3842.5243&-1768.6704&0.2019&0.0515&0.055\\
SC-Storage&Delivery&40&33.25&6462.0&0.0&nS&-4055.77&-4028.059&0.1572&0.1077&0.0065\\
SC-Storage&Delivery&40&33.25&7353.0&0.5&nS&-3697.4575&-3667.3794&0.111&0.0783&0.0066\\
SC-Storage&Delivery&40&33.25&7969.0&1.0&nS&-2875.4661&-2858.7095&0.0686&0.0539&0.0049\\
SC-Storage&Delivery&40&33.25&8344.0&1.5&nIL&-2059.3144&-938.188&0.041&0.5172&-2.1644\\
SC-Storage&Delivery&40&33.25&8555.0&2.0&nIL&-1347.2811&-298.7718&0.0229&0.381&-1.9885\\
SC-Storage&Delivery&40&33.25&8676.0&2.5&nIL&-812.9044&-0.0&0.012&0.4139&-1.5976\\
SC-Storage&Factory Issue&40&33.25&2371.0&-1.5&IL&-485.0832&-386.5032&0.032&&-0.1119\\
SC-Storage&Factory Issue&40&33.25&3741.0&-1.0&nTA&-2383.5457&-2097.5781&0.1612&0.2723&0.7876\\
SC-Storage&Factory Issue&40&33.25&5560.0&-0.5&nS&-3829.363&-3774.8594&0.184&0.0905&0.009\\
SC-Storage&Factory Issue&40&33.25&6828.0&0.0&nS&-3984.1511&-3946.2844&0.1397&0.0856&0.0081\\
SC-Storage&Factory Issue&40&33.25&7596.0&0.5&nS&-3425.5041&-3401.118&0.0944&0.0666&0.0064\\
SC-Storage&Factory Issue&40&33.25&8078.0&1.0&nS&-2732.9244&-2716.7699&0.0628&0.048&0.0054\\
SC-Storage&Factory Issue&40&33.25&8417.0&1.5&nS&-1887.0755&-1876.0928&0.0361&0.0286&0.005\\
SC-Storage&Factory Issue&40&33.25&8579.0&2.0&nS&-1320.8948&-1316.3079&0.0223&0.0193&0.0034\\
SC-Storage&Factory Issue&40&33.25&8679.0&2.5&nS&-838.3075&-836.7675&0.0124&0.0114&0.0021\\
SC-Storage&Sales Order&40&33.25&1563.0&-1.5&nIL&-793.5674&-0.0&0.1126&0.3555&2.435\\
SC-Storage&Sales Order&40&33.25&3712.0&-1.0&nTA&-2818.7441&-2128.7759&0.2196&0.0761&0.2889\\
SC-Storage&Sales Order&40&33.25&5796.0&-0.5&I&-4461.6292&-4461.6292&0.2253&&0.2253\\
SC-Storage&Sales Order&40&33.25&6912.0&0.0&nTA&-4410.7594&-4410.5052&0.1616&0.1616&0.926\\
SC-Storage&Sales Order&40&33.25&7642.0&0.5&nTA&-3727.0848&-3725.165&0.106&0.1063&0.7952\\
SC-Storage&Sales Order&40&33.25&8118.0&1.0&I&-2833.4219&-2833.4219&0.0655&&0.0655\\
SC-Storage&Sales Order&40&33.25&8402.0&1.5&I&-2081.0038&-2081.0038&0.0412&&0.0412\\
SC-Storage&Sales Order&40&33.25&8578.0&2.0&nIL&-1396.5138&-1212.6724&0.0239&0.083&-1.8369\\
SC-Storage&Sales Order&40&33.25&8677.0&2.5&nIL&-900.715&-270.0&0.0136&0.6987&-1.6692\\
SC-Group&Production&40&8.95&2357.0&-2.0&nTA&-280.6608&-0.0635&0.0161&0.0001&0.5215\\
SC-Group&Production&40&8.95&3477.0&-1.5&nS&-1718.6032&-1711.7589&0.1081&0.0723&0.0078\\
SC-Group&Production&40&8.95&5588.0&-1.0&nS&-3471.6838&-3463.4011&0.1546&0.1294&0.0097\\
SC-Group&Production&40&8.95&6714.0&-0.5&nS&-3759.5524&-3731.0882&0.1309&0.1019&0.0165\\
SC-Group&Production&40&8.95&7417.0&0.0&nS&-3367.2243&-3336.8095&0.0953&0.0746&0.0169\\
SC-Group&Production&40&8.95&7941.0&0.5&nS&-2812.1145&-2785.4339&0.0669&0.0543&0.0159\\
SC-Group&Production&40&8.95&8315.0&1.0&nS&-2093.7651&-2073.8308&0.0421&0.0347&0.0159\\
SC-Group&Production&40&8.95&8512.0&1.5&nS&-1499.6157&-1472.4244&0.0264&0.0208&0.02\\
SC-Group&Production&40&8.95&8658.0&2.0&nS&-894.1557&-879.4859&0.0135&0.0111&0.0166\\
SC-Group&Production&40&8.95&8729.0&2.5&nS&-559.9726&-559.9478&0.0076&0.0075&0.0007\\
SC-Group&Delivery&40&8.95&2215.0&-2.0&nTA&-42.2179&-33.4582&0.0018&0.0082&0.7105\\
SC-Group&Delivery&40&8.95&2360.0&-1.5&nTA&-418.985&-26.0499&0.0267&0.0001&0.3116\\
SC-Group&Delivery&40&8.95&3145.0&-1.0&I&-1975.1713&-1975.1713&0.1574&&0.1574\\
SC-Group&Delivery&40&8.95&5295.0&-0.5&nS&-3842.5243&-3830.2557&0.2019&0.171&0.0123\\
SC-Group&Delivery&40&8.95&6462.0&0.0&nS&-4055.77&-4023.5049&0.1572&0.1238&0.0186\\
SC-Group&Delivery&40&8.95&7353.0&0.5&nS&-3697.4575&-3658.958&0.111&0.0873&0.02\\
SC-Group&Delivery&40&8.95&7969.0&1.0&nS&-2875.4661&-2837.9333&0.0686&0.0548&0.0195\\
SC-Group&Delivery&40&8.95&8344.0&1.5&nS&-2059.3144&-2040.8137&0.041&0.035&0.0144\\
SC-Group&Delivery&40&8.95&8555.0&2.0&nS&-1347.2811&-1338.0474&0.0229&0.0203&0.0111\\
SC-Group&Delivery&40&8.95&8676.0&2.5&nS&-812.9044&-805.0809&0.012&0.0106&0.0111\\
SC-Group&Factory Issue&40&8.95&2371.0&-1.5&nTA&-485.0832&-0.0635&0.032&0.0001&0.4022\\
SC-Group&Factory Issue&40&8.95&3741.0&-1.0&nTA&-2383.5457&-491.798&0.1612&0.1479&0.1742\\
SC-Group&Factory Issue&40&8.95&5560.0&-0.5&nS&-3829.363&-3808.9053&0.184&0.1467&0.0153\\
SC-Group&Factory Issue&40&8.95&6828.0&0.0&nTA&-3984.1511&-1365.7829&0.1397&0.1568&0.0203\\
SC-Group&Factory Issue&40&8.95&7596.0&0.5&nS&-3425.5041&-3392.3581&0.0944&0.0762&0.0175\\
SC-Group&Factory Issue&40&8.95&8078.0&1.0&nS&-2732.9244&-2699.4079&0.0628&0.0504&0.0189\\
SC-Group&Factory Issue&40&8.95&8417.0&1.5&nS&-1887.0755&-1860.4191&0.0361&0.0294&0.0182\\
SC-Group&Factory Issue&40&8.95&8579.0&2.0&nS&-1320.8948&-1299.8135&0.0223&0.0184&0.0181\\
SC-Group&Factory Issue&40&8.95&8679.0&2.5&nS&-838.3075&-829.0568&0.0124&0.0109&0.013\\
SC-Group&Sales Order&40&8.95&1563.0&-1.5&nTA&-793.5674&-0.0&0.1126&0.2149&0.1503\\
SC-Group&Sales Order&40&8.95&3712.0&-1.0&I&-2818.7441&-2818.7441&0.2196&&0.2196\\
SC-Group&Sales Order&40&8.95&5796.0&-0.5&nTA&-4461.6292&-4459.0205&0.2253&0.2308&0.9184\\
SC-Group&Sales Order&40&8.95&6912.0&0.0&I&-4410.7594&-4410.7594&0.1616&&0.1616\\
SC-Group&Sales Order&40&8.95&7642.0&0.5&I&-3727.0848&-3727.0848&0.106&&0.106\\
SC-Group&Sales Order&40&8.95&8118.0&1.0&nTR&-2833.4219&-2831.9579&0.0655&0.0652&0.8204\\
SC-Group&Sales Order&40&8.95&8402.0&1.5&nTR&-2081.0038&-2080.3758&0.0412&0.0411&0.7277\\
SC-Group&Sales Order&40&8.95&8578.0&2.0&nS&-1396.5138&-1395.9225&0.0239&0.0234&0.0023\\
SC-Group&Sales Order&40&8.95&8677.0&2.5&nS&-900.715&-895.3006&0.0136&0.0126&0.0083\\
baboons&Affiliative&19&3.648&75.0&0&nIL&-30.1635&-29.9613&0.9199&0.8968&-0.9299\\
baboons&Affiliative&19&3.648&75.0&1&nIL&-30.1635&-29.9613&0.9199&0.8968&-0.9307\\
baboons&Affiliative&19&0.985&75.0&5&nIL&-30.1635&-29.9945&0.9199&0.8847&-2.3127\\
baboons&Affiliative&19&0.598&76.0&10&nIL&-30.2829&-28.822&0.9209&0.8681&-2.4746\\
baboons&Agonistic&19&3.648&595.0&0&nTR&-286.8941&-285.7114&0.1043&0.0998&0.988\\
baboons&Agonistic&19&3.648&595.0&1&nTR&-286.894&-285.5163&0.1042&0.1013&0.9972\\
baboons&Agonistic&19&0.985&596.0&5&nS&-287.0526&-285.2959&0.104&0.0976&0.0549\\
baboons&Agonistic&19&0.598&598.0&10&nS&-287.369&-285.0935&0.1037&0.098&0.0709\\
baboons&Submission&19&3.648&636.0&0&nTR&-161.1799&-157.1451&0.0425&0.0347&0.426\\
baboons&Submission&19&3.648&636.0&1&nTR&-161.1799&-156.6722&0.0425&0.0402&0.3988\\
baboons&Submission&19&0.985&637.0&5&nS&-161.2424&-161.0284&0.0424&0.0413&0.0224\\
baboons&Submission&19&0.598&639.0&10&nS&-161.3672&-161.3322&0.0423&0.0419&0.0119\\
baboons&Attacking&19&3.648&655.0&0&I&-74.6505&-74.6505&0.0153&&0.0153\\
baboons&Attacking&19&3.648&655.0&1&I&-74.6505&-74.6505&0.0153&&0.0153\\
baboons&Attacking&19&0.985&656.0&5&I&-74.6727&-74.6727&0.0153&&0.0153\\
baboons&Attacking&19&0.598&658.0&10&I&-74.7169&-74.7169&0.0152&&0.0152\\
baboons&Chasing&19&3.648&649.0&0&nS&-102.9162&-102.9084&0.0231&0.0229&0.0034\\
baboons&Chasing&19&3.648&649.0&1&nS&-102.9162&-102.9084&0.0231&0.0229&0.0034\\
baboons&Chasing&19&0.985&650.0&5&nS&-102.9499&-102.1694&0.0231&0.022&0.0565\\
baboons&Chasing&19&0.598&652.0&10&I&-103.0171&-103.0171&0.023&&0.023\\
baboons&Embracing&19&3.648&627.0&0&I&-182.3219&-182.3219&0.051&&0.051\\
baboons&Embracing&19&3.648&627.0&1&I&-182.3219&-182.3219&0.051&&0.051\\
baboons&Embracing&19&0.985&628.0&5&I&-182.3974&-182.3974&0.051&&0.051\\
baboons&Embracing&19&0.598&630.0&10&nS&-182.5481&-182.4664&0.0508&0.0503&0.0224\\
baboons&Supplanting&19&3.648&652.0&0&nS&-80.5173&-80.0151&0.0169&0.0153&0.0158\\
baboons&Supplanting&19&3.648&652.0&1&nS&-80.5173&-80.0154&0.0169&0.0153&0.0161\\
baboons&Supplanting&19&0.985&653.0&5&nS&-80.5418&-79.0992&0.0169&0.0147&0.0376\\
baboons&Supplanting&19&0.598&655.0&10&nS&-80.5908&-78.9054&0.0168&0.0145&0.0437\\
baboons&Grooming&19&3.648&446.0&0&nS&-385.7046&-384.5329&0.287&0.2626&0.0262\\
baboons&Grooming&19&3.648&446.0&1&nS&-385.7046&-384.5329&0.287&0.2626&0.0262\\
baboons&Grooming&19&0.985&447.0&5&nS&-387.5015&-384.987&0.2886&0.2689&0.0753\\
baboons&Grooming&19&0.598&448.0&10&nS&-387.9921&-387.6401&0.2879&0.2831&0.0317\\
baboons&Mounting&19&3.648&641.0&0&I&-133.3641&-133.3641&0.0329&&0.0329\\
baboons&Mounting&19&3.648&641.0&1&I&-133.3639&-133.3639&0.0328&&0.0328\\
baboons&Mounting&19&0.985&642.0&5&I&-133.412&-133.412&0.0327&&0.0327\\
baboons&Mounting&19&0.598&644.0&10&I&-133.5078&-133.5078&0.0326&&0.0326\\
baboons&Resting&19&3.648&190.0&0&nIL&-175.1353&-171.7325&0.6631&0.2243&0.0925\\
baboons&Resting&19&3.648&190.0&1&nIL&-175.1353&-171.7325&0.6631&0.2243&0.0925\\
baboons&Resting&19&0.985&191.0&5&nIL&-176.6977&-175.2388&0.6597&0.2769&0.144\\
baboons&Resting&19&0.598&192.0&10&nIL&-177.2959&-173.2492&0.6614&0.2788&0.3785\\
baboons&Playing with&19&3.648&545.0&0&nTA&-295.8915&-265.6715&0.1248&0.0462&0.9992\\
baboons&Playing with&19&3.648&545.0&1&nTA&-295.8915&-265.667&0.1248&0.0462&0.9993\\
baboons&Playing with&19&0.985&545.0&5&I&-295.8915&-295.8915&0.1248&&0.1248\\
baboons&Playing with&19&0.598&547.0&10&I&-296.2753&-296.2753&0.1243&&0.1243\\
baboons&Invisible&19&3.648&390.0&0&nTA&-370.5311&-367.7096&0.3692&0.2851&0.9998\\
baboons&Invisible&19&3.648&390.0&1&nTA&-370.5311&-367.7098&0.3692&0.2851&0.9998\\
baboons&Invisible&19&0.985&391.0&5&nS&-371.1948&-371.0767&0.3683&0.3635&0.0208\\
baboons&Invisible&19&0.598&392.0&10&nS&-371.8564&-371.7842&0.3674&0.3649&0.0192\\
 \hline
\end{longtable}
\end{center}
}

\printbibliography


\end{document}